\documentclass[12pt]{article}
\usepackage{amsmath}
\usepackage[sectionbib]{natbib}
\usepackage{array,epsfig,fancyheadings,rotating}
\usepackage{bm, bbm}
\usepackage{graphicx,subcaption}
\usepackage{multirow}
\usepackage{booktabs}
\usepackage[]{hyperref} 
\usepackage{url} 

\usepackage{amssymb}
\usepackage{amsfonts}
\usepackage{amsthm}
\usepackage[dvipsnames]{xcolor}

\newtheorem{theorem}{Theorem}
\newtheorem{lemma}{Lemma}

\theoremstyle{definition}
\newtheorem{definition}{Definition}

\newtheorem{remark}{Remark}
\newtheorem{assumption}{Assumption}

\newcommand{\bP}{\mathbf{P}}

\def\x{{\bf x}}
\def\y{{\bf y}}
\def\t{{\bf t}}
\def\s{{\bf s}}

\def\A{{\bf A}}
\def\X{{\bf X}}
\def\Y{{\bf Y}}
\def\K{{\bf K}}
\def\L{{\bf L}}

\def\U{{\bf U}}
\def\bS{{\bf S}}
\def\W{{\bf W}}
\def\M{{\bf M}}
\def\R{{\bf R}}

\def\mR{\mathbb{R}}
\renewcommand{\d}{\mathrm{d}}

\def\eeqr{\end{eqnarray}}
\def\beqr{\begin{eqnarray}}
\def\eeqrs{\end{eqnarray*}}
\def\beqrs{\begin{eqnarray*}}

\def\pr{\mathbb{P}}
\newcommand{\defn}{\stackrel{\mbox{{\tiny def}}}{=}}

\newcommand{\blind}{0}

\begin{document}

\def\spacingset#1{\renewcommand{\baselinestretch}%
{#1}\small\normalsize} \spacingset{1}


\if0\blind
{
  \title{\bf RKHS-based Latent Position Random Graph Correlation}
  \author{Xiaoyi Wen\\
     Institute of Statistics and Big Data, Renmin University of China\\
    Junhui Wang\\
    Department of Statistics, The Chinese University of Hong Kong\\
    and \\
    Liping Zhu \\
    Institute of Statistics and Big Data, Renmin University of China}
  \maketitle
} \fi

\if1\blind
{
  \bigskip
  \bigskip
  \bigskip
  \begin{center}
    {\LARGE\bf Title}
\end{center}
  \medskip
} \fi

\bigskip
\begin{abstract}

We study the problem of evaluating the correlation between two latent position random graphs, which is a challenging issue because the latent positions are generally unobservable in random graphs. We introduce a  correlation coefficient that is defined in the reproducing kernel Hilbert space and   merely uses   spectral decomposition of adjacency matrices.   We show that, remarkably, the sample graph covariance converges in probability to its population version even when no kernel function is specified.  
We further design a  permutation  procedure to test the independence between two random graphs.  We assess the performance  of our proposal via simulations and real data analysis, and extend our proposal to spectral decomposition of normalized Laplacian matrices and inhomogeneous random graphs.

\end{abstract}

\noindent%
\vfill

\newpage
\spacingset{1.75} 
\section{Introduction}
\label{sec:intro}

Identifying statistically significant dependencies among data attribute values is crucial for further analysis. Various methods have been developed to test for linear and nonlinear dependence in structured data, yielding satisfactory results. Nonetheless, techniques for detecting dependencies in high-dimensional or complex unstructured data warrant further development.

Analyzing social science, biology, and economics network data presents numerous challenges. In many random graph analyses, identifying correlations between two sets of graphs and testing for independence are crucial steps. Calculating the correlation coefficient between random graphs can provide additional information and facilitate statistical inference. For instance, correlations between protein networks can indicate that proteins with distinct functions participate in the same biological process \citep{ito2001comprehensive}, which is relevant for predicting drug-target binding, side effects, and adverse drug interactions \citep{chu2008construction, kuhn2008large, klipp2010biochemical}. Dysfunction in specific regions of the brain that are part of the brain network associated with drug addiction can lead to habitual abuse. By comparing correlation differences between functional networks of different brain regions in patients and healthy controls, one can identify which areas play a role in drug addiction behavior \citep{janes2012prefrontal, sjoerds2017loss}. Additionally, correlations between brain network activation can reflect the unique synergistic relationship between neurons \citep{dorum2016age}. In social science and business analysis, the correlation analysis of international trade and financial networks can provide new explanations for the spread of financial crises \citep{schiavo2010international}. Furthermore, correlations between stock trading networks on different dates can help reveal stock market inter-dependency \citep{liu2012dynamics}.

Statistical methods for measuring correlation between networks remain limited. Classical coefficients such as Pearson’s capture only linear or monotone relationships and are unsuitable for complex network dependencies, where edges in adjacency matrices are inherently correlated. A common approach is to embed node attributes into low-dimensional vectors and then apply standard correlation measures \citep{lee2019network, xiong2019graph}. Another research line relies on explicit random graph models. \citet{fosdick2015testing} assumed Gaussian latent positions and proposed a likelihood ratio test for dependence between networks and node covariates. For the classical Erd\H{o}s–R'enyi (ER) model, dependence testing is closely related to graph matching, and several recent works have investigated correlated ER graphs using total-variation or spectral-based statistics \citep{ding2023detection, wu2023testing, song2023independence}. Extensions to inhomogeneous ER models have also been explored \citep{sussman2019matched, ding2025efficiently}. These studies primarily focus on detecting the presence of correlation and on establishing detection thresholds or computational limits.

In contrast, our work seeks to develop a measure of graph correlation that satisfies zero–independence equivalence and provides a natural interpretation of correlation strength. Classical independence tests, such as Pearson’s correlation or distance correlation \citep{szekely2007measuring}, are limited in capturing complex nonlinear dependencies. The kernel-based approach \citep{gretton2005measuring} offers a flexible alternative, particularly effective in detecting nonlinear relationships and often exhibit superior finite-sample power compared to distance correlation. In the random graph setting, kernel methods offer a flexible framework that can be viewed in two ways: either as using a kernel function to generate the link probability matrix $\bP$, as in latent position random graphs \citep{hoff2002latent}, or as employing kernel-based statistics for downstream tasks such as two-sample testing \citep{tang2017nonparametric} and community detection \citep{kloster2014heat}.

While kernel methods have shown great promise in various applications, there is still a lack of research on their use in independence testing of random graphs. The flexibility of kernel methods allows for effective analysis of nonlinear dependencies and is also closely related to generating random graphs. Therefore, this paper presents several key contributions:
\begin{itemize}
\item We propose a novel method for testing  independence of random graphs using kernel methods. Our approach does not require explicit specification of the kernel function.
\item We prove that the sample covariance estimated by the spectral decomposition of the adjacency matrix under the latent position graph model converges in probability to its population version.
\item We demonstrate the consistency of the estimated covariances using the normalized Laplacian spectral decomposition and the consistency in inhomogeneous random graphs.
\end{itemize}

The article is structured as follows: Section \ref{sec2} introduces the latent position random graph generation and the relationship between distance correlation and Hilbert-Schmidt correlation to define the population graph correlation ($\mathrm{Gcor}$) and the consistent estimate of  $\mathrm{Gcor}$. In Section \ref{sec3}, the sample $\mathrm{Gcor}$ estimation is achieved by spectral decomposition of the adjacency matrix. We prove its convergence to the population version and demonstrate the validity and consistency of $\mathrm{Gcor}$ using permutation tests. Section \ref{sec4} discusses the relationship between the independence test of random graphs and traditional two-sample tests, demonstrates the consistency of independence tests using normalized Laplacian spectral decomposition, and extends the method to inhomogeneous random graphs. Section \ref{sec5} presents numerical simulation results and actual data applications of the proposed method, and Section \ref{sec6} concludes the article. The on-line Supplementary Material contains all the proofs and the setup for the simulation part.

\section{Independence Test in Random Graph}\label{sec2}

\subsection{Distance Correlation and Hilbert–Schmidt Independence Criterion}
The setting of the hypothesis testing of independence is as follows: given an independent and identically distributed sample $Z=\left \{ (X_{i}, Y_{i})_{i=1}^{n} \right \} \sim F_{XY}$, we want to test
\begin{equation*}
	\begin{aligned}
		& H_{0}:F_{XY}=F_{X}F_{Y}, 
		& H_{1}:F_{XY}\ne F_{X}F_{Y}.
	\end{aligned}
\end{equation*}
We  briefly review distance correlation \citep{szekely2007measuring,szekely2009brownian} and Hilbert–Schmidt independence criterion \citep{gretton2007kernel}.  To test for independence and measure the degree of nonlinear association   between   $\x \in \mathbb{R}^{p}$ and $\y \in \mathbb{R}^{q}$, \cite{szekely2007measuring} define distance covariance by
\begin{equation*}
	\mathrm{dCov}(\x,\y)=\int_{\mathbb{R}^{p+q}}\frac{\left | \phi_{\x,\y}(\t,\s)-\phi_{\x}(\t)\phi_{\y}(\s) \right |^2}{c_{p}c_{q}\left \| \t \right \|_2^{1+p}\left \| \s \right \|_2^{1+q}}(\d \t) (\d \s), 
\end{equation*}
where $\phi(\cdot)$ and $\Gamma(\cdot)$ stand respectively for the characteristic  and gamma functions, $c_{p}=\pi^{(1+p)/2}/\Gamma \{(1+p)/2)\}$ and $\|\cdot\|_2$ stands for the Euclidean norm.  \cite{szekely2007measuring} derive an explicit form of
\begin{align*}
    \mathrm{dCov}(\x,\y) &=\mathrm{E} \left( \left \| \x-{\x}' \right \|_2 \left \| \y-{\y}' \right\|_2 \right) + \mathrm{E}\left( \left\| \x-{\x}' \right\|_2 \right) \mathrm{E} \left( \left\| \y-{\y}' \right\|_2 \right)\\
    & \quad -2\mathrm{E} \left( \left \| \x-{\x}' \right \|_2 \left \| \y-{\y}'' \right\|_2 \right),
\end{align*}
where $({\x}',{\y}')$ and $({\x}'',{\y}'')$ are independent copies of $(\x,\y)$.   \cite{gretton2005measuring} propose to kernelize the Euclidean distance in the reproducing kernel Hilbert space, and define the Hilbert-Schmidt independence criterion (hCov), usually abbreviated as HSIC for simplicity, by
\begin{align*}
    \mathrm{hCov}(\x,\y)&=\mathrm{E}\left \{ K_1(\x,{\x}')K_2(\y,{\y}') \right \}+\mathrm{E}\left \{ K_1(\x,{\x}') \right \}\mathrm{E}\left \{ K_2(\y,{\y}') \right \} \\
    & \quad -2\mathrm{E}\left \{ K_1(\x,{\x}')K_2(\y,{\y}'') \right\},
\end{align*}
where $K_1$ and $K_2$ are user-specified kernels. 

\cite{szekely2014partial} propose a U-centralized procedure to construct an unbiased estimate of distance covariance.  To be precise, define  $\widetilde{B}_{ij}$ to be the U-centralized $B_{ij} \defn  \| \x_i - \x_j\|_2$, that is,
\begin{align*}
    \widetilde{B}_{ij}=B_{ij}-\frac{1}{(n-2)}\sum_{j=1}^{n}B_{ij}-\frac{1}{(n-2)}\sum_{i=1}^{n}B_{ij}+\frac{1}{(n-1)(n-2)}\sum_{i,j=1}^{n}B_{ij}, 
\end{align*}
for $i \ne j$ and 0 for $i=j$.
Throughout we use the notation $\widetilde{B}_{ij}$ as a U-centralized operator of $B_{ij}$.
Similarly, we define $ \widetilde{C}_{ij}$ and $  \widetilde{R}_{ij}$ and $  \widetilde{H}_{ij}$ to be the respective U-centralized $C_{ij} \defn  \| \y_i - \y_j\|_2$,  $R_{ij}  \defn  K_1( \x_i, \x_j)$ and $H_{ij} \defn  K_2( \y_i, \y_j)$. The unbiased U-estimates of $\mathrm{dCov}(\x,\y) $ and $\mathrm{hCov}(\x,\y)$ are given respectively by
  \beqrs
 	\mathrm{dCov}_{n}(\x,\y) \defn \frac{1}{n(n-3)}\sum_{i\neq j}\widetilde{B}_{ij}\widetilde{C}_{ij}, \textrm{\  and \  }
 	\mathrm{hCov}_{n}(\x,\y) \defn \frac{1}{n(n-3)}\sum_{i\neq j}\widetilde{R}_{ij}\widetilde{H}_{ij}.
 \eeqrs

\subsection{Random Graph Correlation}
Let us first define the independence between the latent position random graphs, $G_1$ and $G_2$, that have exactly the same vertex set $\{1,\ldots, n\}$. This depends on how these graphs are structured.   We assume that $G_1$ and $G_2$ are structured with latent positions. To be precise,   let  $\{\x_1,\ldots, \x_n\}$ and  $\{\y_1,\ldots, \y_n\}$  be the respective latent positions  of $G_1$ and $G_2$.    We further assume that  $\{\x_1,\ldots, \x_n\}$ are independent and follow the identical distribution $F_X$, and $\{\y_1,\ldots, \y_n\}$ are independent and follow the identical distribution $F_Y$. In symbols, $\x$ follows $F_X$ and $\y$ follows $F_Y$.  Let $F_{XY}$ be the joint distribution of   $(\x,\y)$.   We define    $G_1$ and $G_2$  to be independent if and only if  the  latent positions of these random graphs, $\x$ and $\y$, are   independent. This allows us to test the independence between $G_1$ and $G_2$  through testing $H_{0}:F_{XY}=F_{X}F_{Y}$ versus $H_{1}:F_{XY}\ne F_{X}F_{Y}$.

Next let us explore how to measure the  strength of dependence between $G_1$ and $G_2$. Denote the adjacency matrices of  $G_1$ and $G_2$ by $\A^{G_1} = (A^{G_1}_{ij})_{n\times n}$ and $\A^{G_2} = (A_{ij}^{G_1})_{n\times n}$, respectively.  \cite{lyzinski2014seeded}  proposed to measure the degree of dependence between $G_1$ and $G_2$  by  calculating the correlation between $\{(A_{ij}^{G_1}, A_{ij}^{G_2}), 1\le i  < j\le n\}$ directly, which completely ignores the structures of random graphs. The latent position random  graph   $G_1$ is said to be generated by the kernel function $K$, if $A_{ij}^{G_1}$ follows a  Bernoulli distribution such that $\pr(A_{ij}^{G_1} = 1)  = K_{ij} = 1- \pr(A_{ij}^{G_1} = 0)$,   where $K_{ij} \defn  K(\x_i, \x_j)$. Similarly, the latent position random  graph  $G_2$ is said to be generated by the kernel function $L$, if $A_{ij}^{G_2}$   follows a  Bernoulli distribution such that  $\pr(A_{ij}^{G_2} = 1)  = L_{ij} = 1- \pr(A_{ij}^{G_2} = 0)$,   where $L_{ij} \defn  L(\y_i, \y_j)$.  We define  $K_{ii} = L_{ii} = 0$, for $i=1,\ldots, n$, to exclude self-loops.  In addition, a kernel is said to be  universal if it encodes all  information  of the distribution \citep{fukumizu2007kernel}.

For notation clarity, we write $\X \defn (\x_1,\ldots,\x_n)^\top$, $\Y \defn (\y_1,\ldots,\y_n)^\top$,    $\mathbf{K} \defn (K_{ij})_{n\times n}$   and $\mathbf{L} \defn (L_{ij})_{n\times n}$.   
We define the correlation between the latent position random graphs below.
\begin{definition}[Population Graph Correlation]  Suppose  the kernels $K$ and $L$ are  universal. Let $(\x,\y)$, $({\x}',{\y}')$ and $({\x}'',{\y}'')$ be the independent copies. We define the population  graph covariance by
\begin{align}\label{pop_gcov}
    \mathrm{gCov}(G_{1},G_{2}) &\defn  \mathrm{E} \left \{ K(\x,{\x}')L(\y,{\y}') \right \}  +\mathrm{E}\left \{  K(\x,{\x}') \right \}  \mathrm{E} \left \{  L(\y,{\y}') \right \} \nonumber \\
    & \quad -2\mathrm{E} \left \{ K(\x,{\x}')L(\y,{\y}'') \right \}.  
\end{align}
Define the graph variances of $G_1$ and $G_2$ by  $\mathrm{gVar}(G_{1})\defn \mathrm{gCov}(G_{1},G_{1})=\mathrm{E} \left \{ K(\x,{\x}') \right \}^2 +\mathrm{E}^2\left \{  K(\x,{\x}') \right \}    -2\mathrm{E} \left \{ K(\x,{\x}')K(\x,{\x}'') \right \}$ and similarly, $\mathrm{gVar}(G_{2})\defn \mathrm{gCov}(G_{2},G_{2})=\mathrm{E} \left \{ L(\y,{\y}') \right \}^2 +\mathrm{E}^2\left \{  L(\y,{\y}') \right \}    -2\mathrm{E} \left \{ L(\y,{\y}')L(\y,{\y}'') \right \}$. We further define the graph correlation between $G_1$ and $G_2$ by 
\beqr\label{pop_gcor} \mathrm{gCor}(G_{1},G_{2})  &\defn&  \mathrm{gCov}(G_{1},G_{2})/\left \{  \mathrm{gVar}(G_{1}) \mathrm{gVar}(G_{2}) \right \}^{1/2}.
\eeqr
\end{definition}

By definition, the graph covariance and accordingly the graph correlations are zero if and only if the latent positions $\x$ and $\y$ are independent. Therefore, the graph correlation serves as  a valid metric to measure the degree of nonlinear dependence between two random graphs.  We further make two remarks here before we proceed. First,  we use kernel distances to connect the graph correlation with the generation mechanism of  latent position random graphs.  Such graph models are  more general than the random dot product graph model  \citep{young2007random}, and  closely related to the inhomogeneous and exchangeable random graph models \citep{bollobas2007phase,diaconis2007graph}. Therefore, the graph correlation can be readily used in the inhomogeneous random graphs.  Second,   our proposed graph correlation is very different from the Hilbert-Schmidt independence criterion \citep{gretton2005measuring} in that the latent positions are not observable and have to be estimated with graph embedding.  In the sequel we shall see that, the kernels of the graph correlation will be approximated in a data driven fashion. Thus we are not required to select the kernels delicately.

\section{Sample Graph Correlations}\label{sec3}
In this section we propose to estimate the graph correlations at the sample level. The overall estimation procedure can be divided into two steps. In the first step  we approximate the feature maps induced by kernel functions  through  eigen-decomposition of the adjacency matrix in the latent position random graphs. In the second step we estimate the graph correlations  through the truncated feature maps.

\subsection{Approximation of Feature Maps}

In this section, we follow the feature approximation framework proposed by \citet{tang2013universally}, which establishes the consistency of latent position random graphs generated by a kernel function. Building on this idea, we extend the consistency between feature maps to the consistency of the estimator of \eqref{pop_gcov}. Let $\nu$ denote a strictly positive measure on the  Borel $\sigma$-field of $\mathcal{X}$. We further assume that, the kernel function, denoted with $K:\mathcal{X} \times \mathcal{X}\rightarrow [0,1]$, is  continuous and positive definite. We denote the space of square integrable functions on $\nu$ as ${\cal L}^2(\mathcal{X},\nu)$, and define an integral operator,  $T_{K}:{\cal L}^2(\mathcal{X},\nu) \rightarrow {\cal L}^2(\mathcal{X},\nu)$,  by 
\begin{align}\label{eq:def_Tk}
    (T_{K}f) (x) &\defn \int _{\mathcal{X}}K(x,{x}')f({x}')d\nu ({x}'), \textrm{   for  }  f \in {\cal L}^2(\mathcal{X},\nu),
\end{align}
which is  continuous and compact. Let $\{\lambda_i, i = 1, \ldots, \}$  be the eigenvalues of $T_{K}$, and $\left \{ \psi_i, i=1, \ldots,   \right \}$ be the associated orthonormal eigen-functions.  
Through out the paper, we suppose the following assumption holds:
\begin{assumption}\label{assum:assump_Tk}

Let the eigenvalues $(\lambda_i)_{i \in I}$ be simple and positive such that $\lambda_1 > \lambda_2 > \cdots > 0$.
    
\end{assumption}

Suppose $\mathcal{H}$ is the reproducing kernel  Hilbert space induced by $K$. Let $\left \langle\cdot,\cdot  \right \rangle$ be the inner product, and $\bm{\Phi}= (\phi_1, \phi_2,\ldots )$ be the feature map which satisfies  $  \left \langle \bm{\Phi}(x),\bm{\Phi}({x}')  \right \rangle_{\mathcal{H}}=K(x,{x}')$, for all $x,  x' \in \mathcal{X}$. By  the Mercer's representation theorem \citep{cucker2002mathematical}, the feature map
satisfies $\phi_i = \lambda_i^{1/2}\psi_i$,  for $i=1,\ldots$. Define $\bm{\Phi}_d$ to be a truncation of $\bm{\Phi}$, that is,  $\bm{\Phi}_d \defn (\phi_1,\ldots, \phi_d)$.

We define the adjacency spectral embedding as follows. Suppose a random graph $G = (V, E)$ is generated by latent positions $\X \in \mathbb{R}^{n \times p}$ and has an adjacency matrix $\A \in \mR^{n \times n}$.  We impose the  spectral  decomposition onto the adjacency matrix to obtain  $ (\A^\top \A)^{1/2} =  \U\bS  \U^\top$,  where   $\bS\in\mR^{n\times n}$ is a diagonal matrix consisting of the descending eigenvalues of $(\A^\top \A)^{1/2}$, and $\U\in\mR^{n\times n}$ is an ortho-normal matrix consisting of the eigenvectors associated with the corresponding eigenvalues. We further define $\bS_d\in\mR^{d\times d}$ and $\U_d\in\mR^{n\times d}$ to be the respective $d$-truncation of  $\bS$ and $\U$. In other words,  $\bS_d$  is composed by the first $d$  rows and columns of $\bS$, and $\U_d$ is composed by the first $d$ columns of  $\U$.  Following \citep{fishkind2013consistent}, we define the  adjacency spectral embedding of $(\A^\top \A)^{1/2}$ into $\mR^d$  by $\U_d\bS_d^{1/2}$. Using the estimated feature map we can estimate the kernel matrix even when both the kernel function and latent positions are unknown. It is worth noting that the consistency results in \cite{tang2013universally} are established for relatively dense graphs, therefore we introduce the following assumption.
\begin{assumption}\label{assum:graph_density}

Given a fixed sequence $(\x_i)_{i=1}^n$, let $\Delta \defn \max_{i} \sum_{j=1}^n K(\x_i,\x_j)$. There exists a universal constant $c_0 > 0$ such that $\Delta > c_0 \log(n)$ holds. 
\end{assumption}

\begin{remark}
    Assumption~\ref{assum:assump_Tk} requires the eigenvalues of the kernel integral operator to be distinct, a condition that is standard in the kernel–PCA literature \citep{rudi2015less, sterge2020gain}. Moreover, \citet[Proposition 1]{shi2009data} shows that this assumption holds analytically for Gaussian kernels with Gaussian distributions. Assumption~\ref{assum:concentration_sparse} imposes a condition on the graph density to guarantee the concentration property of dense random graphs \citep{oliveira2009concentration}.
\end{remark}


\subsection{Definition}

The estimation process of the population graph correlation is as follows. For the latent position random graphs $G_{1}$ generated by the latent position matrix $\X \in \mathbb{R}^{n \times p}$ with kernel function $K,$ the kernel matrix $\mathbf{K}=(K_{ij})_{n \times n}$ with entries $K_{ij}=K(\x_i,\x_j)$ for $i \ne j$ and 0 otherwise. Similarly, for $G_{2}$ are generated by the latent position matrix $\Y \in \mathbb{R}^{n \times q}$ with kernel function $L$, we have kernel matrix $\mathbf{L}=(L_{ij})_{n \times n}$ where $L_{ij}=L(\y_i,\y_j)$ for $i \ne j$ and 0 otherwise. According to Mercer's representation theorem, the $d$-th truncated feature map for $\mathbf{K}$ and $\mathbf{L}$ are denoted by $\bm{\Phi}_{1,d}$ and $\bm{\Phi}_{2,d}$. Let $\A_{1}$ and $\A_{2}$ be the adjacency matrices of $G_1$ and $G_2$, respectively. We denote $\U_{1}\bS_{1}\U_{1}^{\top}$ and $\U_{2}\bS_{2}\U_{2}^{\top}$ as the sepctral decomposition of $\A_{1}$ and $\A_{2}$. Following the definition of adjacency spectral embedding, let $\bS_{l,d}$ be a diagonal matrix composed of the top $d$ eigenvalues, and $\U_{l,d}$ is the matrix composed by corresponding eigenvectors for $l=1,2$. Denote $\U_{1,d}\bS_{1,d}^{1/2} \in \mathbb{R}^{n \times d}$ by $\hat{\bm{\Phi}}_{1,d}$ and $\U_{2,d}\bS_{2,d}^{1/2} \in \mathbb{R}^{n \times d}$ by $\hat{\bm{\Phi}}_{2,d}$. Then we have the estimated kernel functions $\hat{\mathbf{K}}=\hat{\bm{\Phi}}_{1,d}\hat{\bm{\Phi}}_{1,d}^{\top} \in \mathbb{R}^{n \times n}$ and $\hat{\mathbf{L}}=\hat{\bm{\Phi}}_{2,d}\hat{\bm{\Phi}}_{2,d}^{\top} \in \mathbb{R}^{n \times n}$. The $U$-centralized matrices of the $\hat{\mathbf{K}} = (\hat{K}_{ij})_{n \times n}$ and $\hat{\mathbf{L}} = (\hat{L}_{ij})_{n \times n}$ are defined as $\widetilde{\mathbf{K}}=(\tilde{K}_{ij})_{n \times n}$ and $\widetilde{\mathbf{L}}=(\tilde{L}_{ij})_{n \times n}$, and can be computed by the following formula:
\begin{align*}
    \tilde{K}_{ij} = \hat{K}_{ij}-\frac{1}{n-2}\sum_{j=1}^{n}\hat{K}_{ij}-\frac{1}{n-2}\sum_{i=1}^{n}\hat{K}_{ij}+\frac{1}{(n-1)(n-2)}\sum_{i,j=1}^{n}\hat{K}_{ij} 
\end{align*}
for $i \ne j$ and 0 for $i=j$. Similarly, we can obtain $\widetilde{\mathbf{L}}$. The estimator of population graph covariance can be represented as
\begin{equation}\label{sample_gcov}
    \mathrm{gCov}_{n}(G_{1},G_{2}) = \frac{1}{n(n-3)}\sum_{i\neq j}\tilde{K}_{ij}\tilde{L}_{ij}.
\end{equation}
The sample graph variance and correlations also can be defined as follows:
\begin{gather}\label{sample_gcor}
    \mathrm{gVar}_{n}(G_{1}) = \frac{1}{n(n-3)}\sum_{i\neq j}\tilde{K}_{ij}^2,\nonumber \quad  \mathrm{gVar}_{n}(G_{2}) = \frac{1}{n(n-3)}\sum_{i\neq j}\tilde{L}_{ij}^2,\nonumber \\
    \mathrm{gCor}_{n}(G_{1},G_{2})=\mathrm{gCov}_{n}(G_{1},G_{2})/\left \{ \mathrm{gVar}_{n}(G_{1})\mathrm{gVar}_{n}(G_{2}) \right \}^{1/2} .
\end{gather}

Through the estimation procedure mentioned above, we can acquire a consistent estimator for the proposed graph covariance \textbf{without} needing to specify the kernel function. It should be emphasized that in our estimation process, the choice of embedding dimension $d$ is crucial because we use the spectral decomposition of the adjacency matrix to approximate the feature map of the kernel.


\subsection{Main Results}

By using the spectral adjacency decomposition we can obtain a reliable and consistent estimation of the truncated feature maps of the kernel matrix. As such, the kernel distance matrix evaluated through the spectral decomposition of $(\A^\top \A)^{1/2}$ is an excellent approximation of $\mathbf{K}$. Our proposed sample covariances, variances, and correlations converge to their respective population analogs with increasing node count, and this convergence occurs with high probability. 
\begin{theorem}\label{theorem1}
    Suppose LPRG $G_{1}$ is generated by latent position $\X \in \mathbb{R}^{n \times p}$ and universal kernel $K$, LPRG $G_{2}$ is generated by latent position $\Y \in \mathbb{R}^{n \times q}$ and universal kernel $L$. 
    Let the eigenvalues of integral operators $T_K$ and $T_L$ (defined as \eqref{eq:def_Tk}) are denoted as $\left(\lambda_i (K) \right)_{i \in I}$ and $\left(\lambda_i (L) \right)_{i \in I}$, respectively.
    The population Hilbert-Schmidt covariance is defined as \eqref{pop_gcov}. The sample graph covariance is calculated by \eqref{sample_gcov}. 
    Suppose Assumptions~\ref{assum:assump_Tk} and \ref{assum:graph_density} hold. Thus we have
    \begin{itemize}
        \item Suppose the eigenvalues $(\lambda_i(K))_{i \in I}$ and $(\lambda_i(L))_{i \in I}$ decay polynomially, satisfying $\lambda_i(K) = \mathcal{O}(i^{-\alpha_K})$ and $\lambda_i(L) = \mathcal{O}(i^{-\alpha_L})$ for $\alpha_K, \alpha_L > 1$. Then, there exists a constant $c > 0$ such that for all $d$ satisfying $\max \left\{ n^{1/\alpha^2_K}, n^{1/\alpha^2_L} \right\} \le d \le c n / \log n $, the sample graph covariance $\mathrm{gCov}_n(G_1, G_2)$ converges in probability to $\mathrm{gCov}(G_1, G_2)$ as $n \to \infty$.
        \item Suppose the eigenvalues $(\lambda_i(K))_{i \in I}$ and $(\lambda_i(L))_{i \in I}$ decay exponentially, satisfying $\lambda_i(K) = \mathcal{O}(e^{-\tau_K i})$ and $\lambda_i(L) = \mathcal{O}(e^{-\tau_L i})$ for $\tau_K, \tau_L > 0$. Then, there exists constant $\theta, c > 0$ such that for all $d$ satisfying $\max \left\{ (\theta \log n)/\tau_K , (\theta \log n)/\tau_L \right\} \le d \le c n / \log n $, the sample graph covariance $\mathrm{gCov}_n(G_1, G_2)$ converges in probability to $\mathrm{gCov}(G_1, G_2)$ as $n \to \infty$.
    \end{itemize}

\end{theorem}

The above theorem demonstrates the consistency of our proposed estimator, and the next theorem provides its asymptotic performance under both the null and alternative hypotheses.
\begin{theorem}\label{theorem2}
    Let $G_1$ and $G_2$ be two latent position random graphs generated by universal kernels $K$ and $L$, respectively. Suppose Assumptions~\ref{assum:assump_Tk} and \ref{assum:graph_density} hold. Under the consistency conditions outlined in Theorem~\ref{theorem1}, the following hold:
    \begin{itemize}
        \item $\mathrm{gCor}_n(G_1, G_2) \to 0$ if $G_1$ and $G_2$ are independent.
        \item $\mathrm{gCor}_{n}(G_{1},G_{2}) \rightarrow$ a positive constant when $G_{1}$ and $G_{2}$ are not independent.
    \end{itemize}
\end{theorem}

\begin{remark}
Theorem~\ref{theorem2} establishes the zero–independence equivalence of the $\mathrm{gCor}_n$, it converges to zero if and only if the two graphs are independent. Moreover, its limiting value $\mathrm{gCor}$ is
(i) invariant to graph sparsity, and
(ii) equal to $1$ if and only if the two feature spaces are isometric up to orthogonal transformation and scaling (the RKHS analogue of $Y = \mathbf{a} + b X \mathbf{C}$ in distance correlation \citep[Theorem 3]{szekely2007measuring}). 
\end{remark}

\begin{remark}[Gaussian benchmark and magnitude interpretation]
Consider a parametric benchmark in which the latent positions are jointly Gaussian and a Gaussian kernel is used. Specifically, let $k(x,y) = \exp\!\left(-\frac{(x-y)^2}{2\sigma^2}\right)$ with $\sigma^2=1$ and assume $(X,Y)$ follows a bivariate normal distribution with $X \sim \mathcal{N}(0,1)$, $Y \sim \mathcal{N}(0,1)$, and $\mathrm{Cov}(X,Y)=\rho$. According to the analytical expression of HSIC for this setting \citep[Lemma~1]{kalinke2024minimax}, the population graph covariance satisfies
\[
\mathrm{gCov}^2(G_1,G_2)
= \frac{1}{\sqrt{4-9\rho^2}} + \frac{1}{3}
- \frac{2}{\sqrt{4-\rho^2}},
\]
which is a strictly increasing function of $\rho^2$ for $|\rho|<1$.
Since the normalization terms in $\mathrm{gCor}$ depend only on the marginal distributions of $X$ and $Y$ and are constant with respect to $\rho$, it follows that $\mathrm{gCor}$ is also monotone in $|\rho|$ in this regime. Thus, in the Gaussian setting, larger values of $\mathrm{gCor}$ correspond to stronger latent dependence.
\end{remark}

\subsection{Permutation Test}

We use the permutation test to test the independence of a pair of random graphs \citep{nichols2002nonparametric}. 
The truncated feature maps estimated by eigendecomposition of the adjacency matrices are denoted as $\hat{\bm{\Phi}}_{1,d}$ and $\hat{\bm{\Phi}}_{2,d}$. The initial sample $\mathrm{gCor}$ statistic is calculated by $\hat{\bm{\Phi}}_{1,d}$ and $\hat{\bm{\Phi}}_{2,d}$ and denoted by $\mathrm{gCor}_{n}^{0}(G_{1},G_{2})$. Then shuffle the order of $\hat{\bm{\Phi}}_{2,d}$, record it as $\hat{\bm{\Phi}}_{2,d}^{*}$, calculate the sample $\mathrm{gCor}$ and record it as $\mathrm{gCor}_{n}^{*}(G_{1},G_{2})$. The permutation step is repeated $r$ times, and the following probability obtains the p-value estimate
\begin{equation*}
    \mathbb{P}\left \{ \mathrm{gCor}_{n}^{*}(G_{1},G_{2})>\mathrm{gCor}_{n}^{0}(G_{1},G_{2}) \right \}.
\end{equation*}
The null hypothesis is rejected if the calculated p-value is less than a pre-specified critical level, usually set to 0.05 or 0.01. The following theorem states that the p-value obtained by the permutation test method is valid for testing the independence of random graphs using $\mathrm{gCor}$.
\begin{theorem}\label{theorem5}
    Suppose two latent position random graphs $G_{1}$ and $G_{2}$ are generated by latent variables $\X$ and $\Y$, respectively. Given the Type \uppercase\expandafter{\romannumeral1} error level $\alpha \in (0,1)$, the permutation test based on $\mathrm{gCor}_n$ satisfies: (i) it is asymptotically valid at level $\alpha$ under $H_0$; (ii) it is universally consistent against fixed alternatives.
\end{theorem}

\section{Extensions}\label{sec4}

\subsection{Extension to the normalized Laplacian}\label{sec4_2}

We note that the eigendecomposition of the Laplacian or normalized Laplacian matrices can provide a way to derive feature representations for individual nodes within a random graph. Specifically, the Laplacian matrix, a discrete form of the Laplace-Beltrami operator, has been employed in manifold learning and nonlinear dimension reduction methods, including Laplacian eigenmaps \citep{belkin2003laplacian} and diffusion maps \citep{coifman2006diffusion}. These approaches offer a means to extract meaningful structural information from the graph that can aid in subsequent analyses.

For the matrix $\M$ with non-negative entries, the normalized Laplacian of matrix $\M$ is defined by
\begin{equation}\label{eq:cal_laplacian}
    \mathbf{L}(\M) \defn  \{\mathrm{diag}(\M \mathbf{1})\}^{-1/2}\M \{\mathrm{diag}(\M \mathbf{1})\}^{-1/2}.
\end{equation}
We use the definition of the normalized Laplacian matrix in \cite{tang2018limit}, which may be slightly different from the normalized Laplacian in some papers, such as denoting it as $\mathbf{I}-\mathbf{L}(\M) $ \citep{belkin2003laplacian}. For the graph embedding, which uses the eigenvalues and eigenvectors of the normalized Laplacian, the definitions of these two normalized Laplacian are equivalent. Next, we introduce the definition of Laplacian spectral embedding in random graphs. 

\begin{definition}[Laplacian spectral embedding (LSE)]
Suppose a random graph generated by latent positions $\X \in \mathbb{R}^{n \times p}$ has an adjacency matrix $\A \in \mathbb{R}^{n \times n}$. The normalized Laplacian matrix is denoted as $\mathbf{L}(\A) \in \mathbb{R}^{n \times n}$ calculated by \eqref{eq:cal_laplacian}. Similar to the eigendecomposition of the adjacency matrix, the Laplacian spectral embedding of $\mathbf{L}(\A)$ into $\mathbb{R}^{d}$ is $\tilde{\U}_{A,d}\tilde{\bS}_{A,d}^{1/2}$, where $\tilde{\bS}_{A,d} \in \mathbb{R}^{d \times d}$ is a diagonal matrix composed by the $d$ largest eigenvalues of $\mathbf{L}(\A)$ and $\tilde{\U}_{A,d} \in \mathbb{R}^{n \times d}$ are corresponding eigenvectors.

\end{definition}



The following theorem shows that, although the spectral decomposition of $\mathbf{L}(\A)$ cannot be directly used to approximate the population graph covariance \eqref{pop_gcov}, there exists a consistency relationship between the distance correlation calculated by the spectral embeddings of $\mathbf{L}(\A)$ and $\mathbf{L}(\mathbf{K})$.

\begin{theorem}\label{theorem7}
    Suppose two latent position random graphs $G_{1}$ and $G_{2}$ are generated by latent matrices $\X \in \mathbb{R}^{n \times p}$ and $\Y \in \mathbb{R}^{n \times q}$, respectively, and the kernel matrices are $\K_{1} \in \mathbb{R}^{n \times n}$ and $\K_{2} \in \mathbb{R}^{n \times n}$. Let $\A_{1} \in \mathbb{R}^{n \times n}$ be the adjacency matrix of $G_{1}$ and $\A_{2} \in \mathbb{R}^{n \times n}$ be the adjacency matrix of $G_{2}$. The spectral embedding of $\mathbf{L}(\A_{1})$ and $\mathbf{L}(\A_{2})$ are $\hat{\bm{\Psi}}_{1,d} \in \mathbb{R}^{n \times d}$ and  $\hat{\bm{\Psi}}_{2,d} \in \mathbb{R}^{n \times d}$, respectively. The spectral embedding of $\mathbf{L}(\K_{1})$ and $\mathbf{L}(\K_{2})$ are $\bm{\Psi}_{1,d} \in \mathbb{R}^{n \times d}$  and $\bm{\Psi}_{2,d} \in \mathbb{R}^{n \times d}$, respectively. Then  $\mathrm{dCov}_{n}(\hat{\bm{\Psi}}_{1,d},\hat{\bm{\Psi}}_{2,d}) -  \mathrm{dCov}_{n}(\bm{\Psi}_{1,d},\bm{\Psi}_{2,d})$ converges  in  probability  to  0, as $n\to\infty$,
     where $\mathrm{dCov}_{n}$ denotes the sample version of distance covariance.
\end{theorem}

\begin{remark}

\begin{enumerate}
    \item It is worth noting that the conclusion of Theorem~\ref{theorem7} is based on distance covariance to avoid potential confusion with the kernel function used to generate the random graphs. However, this result also holds for the Hilbert-Schmidt covariance when specific kernel functions are defined.
    \item Theorem~\ref{theorem7} describes a two-step approach for obtaining the graph covariance, aligning with the estimation method presented in \cite{lee2019network}. Our theorem provides theoretical support for this two-step estimation procedure.
\end{enumerate}
    
\end{remark}

\subsection{Extension to Inhomogeneous Random Graphs}\label{sec4_3}
In this section, we extend these conclusions to inhomogeneous random graphs \citep{bollobas2007phase}. Given the independent and identically distributed latent variables $\left\{ \x_{i} \right \}_{i=1}^n \sim F$, the kernel matrix $\K_{n}=\left[ \rho_{n} K(\x_{i},\x_{j}) \right]_{i,j=1}^n$. The $\rho_{n} \in (0,1)$ is a scaling parameter that makes the random graph sparse, which is more reliable with some practical application scenarios. A common choice for $\rho_{n}$ is $\rho_{n}=(\log{n})/n $ \citep{bollobas2007phase, oliveira2009concentration}. 
To relax the density condition stated in Assumption~\ref{assum:graph_density}, we introduce the following two assumptions to ensure the consistency of the proposed $\mathrm{gCov}_n$ in sparse settings.
\begin{assumption}\label{assum:graph_sparsity}
      The scaling parameter $\rho_{n}$ satisfies $n \rho_{n}=\omega \left \{ (\log n)^{1/2+\epsilon} \right\} $ for $\epsilon > 0$\footnote{The notation $a_{n}=\omega (b_{n})$ means for any positive constant $C$, there exists $n_{0}$ such that $a_{n}>C b_{n}$ for all $n \ge n_{0}$.}.
\end{assumption}

\begin{assumption}\label{assum:concentration_sparse}

Let $\A$ be the adjacency matrix of the inhomogeneous random graph and $E(\A )$ denotes its expectation. 
Let $\| \cdot\|$ be a spectral norm. 
Then we have $\left \|  \A - E(\A) \right \| \le (\log n)^{1/2}$ with high probability.

\end{assumption}

\begin{remark}
    Assumption~\ref{assum:graph_sparsity} implicitly imposes a semi-sparse structure on the inhomogeneous random graph under consideration. This semi-sparse graph retains concentration properties, albeit with varying deviation bounds, ensuring that Assumption~\ref{assum:concentration_sparse} is readily satisfied. 
\end{remark}

Consequently, we establish the following theorem for the consistency of the semi-sparse random graph.

\begin{theorem}\label{theorem8}
    Suppose two inhomogeneous random graphs $G_{1}$ and $G_{2}$ are generated by latent positions $\X \in \mathbb{R}^{n \times p}$ and $\Y \in \mathbb{R}^{n \times q}$, respectively. The associated kernels are $K$ and $L$ with the scaling parameter $\rho_{n}$.  
    The population Hilbert-Schmidt covariance is defined as \ref{pop_gcov}. The sample graph covariance is calculated by \eqref{sample_gcov}. 
    Suppose Assumption~\ref{assum:assump_Tk}, \ref{assum:graph_sparsity} and \ref{assum:concentration_sparse} hold.
    Then under the consistency conditions outlined in Theorem~\ref{theorem1} and $d \le c_1 \left( \log n \right)^{2 \epsilon}$ for some constant $c_1 >0$, $\mathrm{gCov}_{n}(G_{1},G_{2}) $ converges in probability to $\mathrm{gCov}(G_{1},G_{2}) $ as $n\to\infty$.
    
\end{theorem}

\section{Numerical Studies}\label{sec5}

\subsection{Simulations}

In this section we provide empirical validation of the proposed test statistic. Our experiments are designed as follows:
\begin{itemize}
    \item \textbf{Study 1 (Convergence performance)}: We demonstrate the convergence performance of the proposed sample $\mathrm{gCorr}$ statistic to its population-level counterpart, the HSIC. Specifically, we examine two linear relationships between latent positions and analyze the impact of noise disturbances in the distribution on the convergence behavior.
    \item \textbf{Study 2 (Comparison with testing power and runtime)}: We compare the testing power and runtime of our approach with the two-step correlation identification method proposed by \cite{lee2019network}, a primary framework for analyzing random graphs. This method first embeds the random graph to obtain low-dimensional vector representations of nodes and then computes the correlation coefficient between these vectors, using techniques such as MGC \citep{shen2020distance} and Dcorr \citep{szekely2007measuring}.
\end{itemize}
 
\begin{figure}
     \centering
     \begin{subfigure}{0.28\textwidth}
         \centering
         \includegraphics[width=\textwidth]{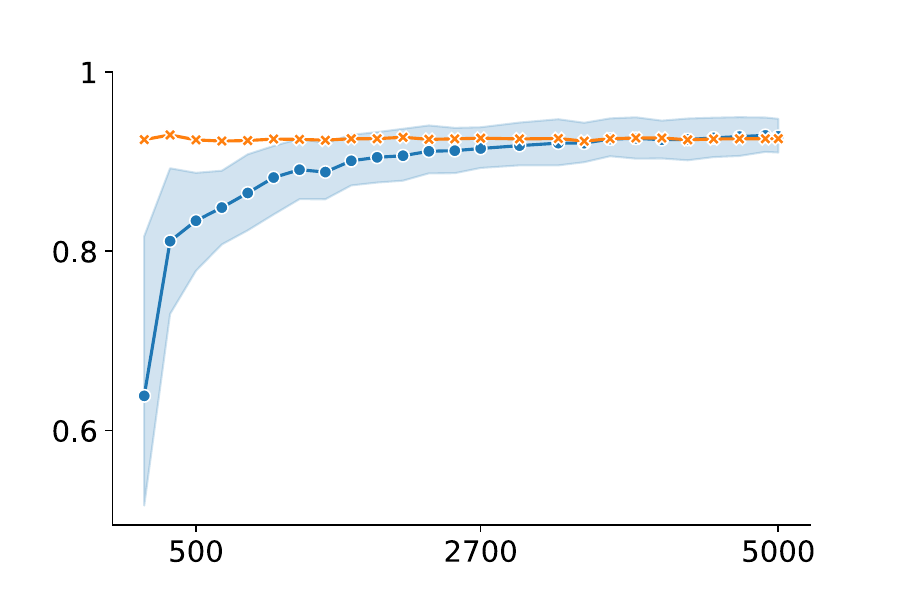}
         \caption{Linear ($\kappa=0$)}
     \end{subfigure}
     \hspace{0.2cm}
     \begin{subfigure}{0.28\textwidth}
         \centering
         \includegraphics[width=\textwidth]{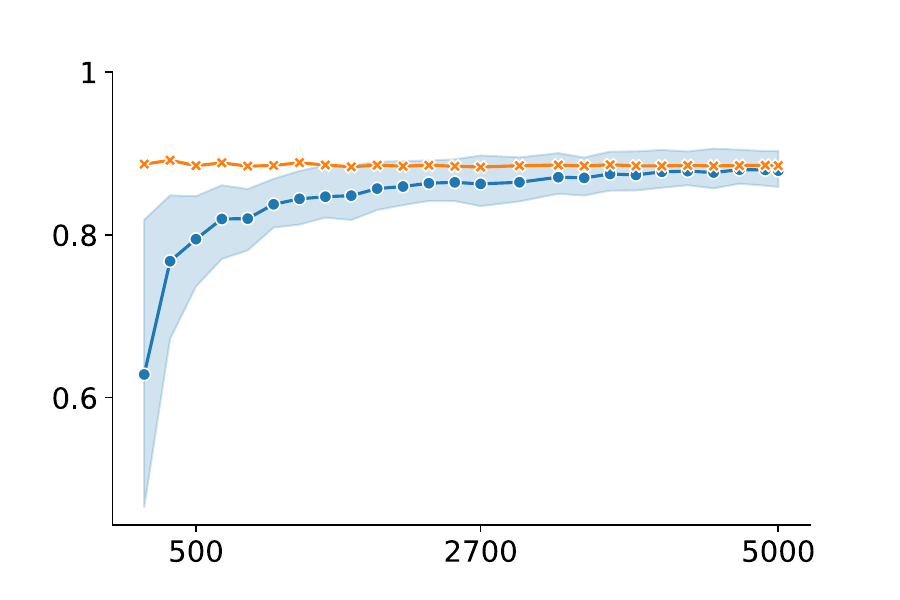}
         \caption{Linear ($\kappa=0.05$)}
     \end{subfigure}
     \hspace{0.2cm}
     \begin{subfigure}{0.28\textwidth}
         \centering
         \includegraphics[width=\textwidth]{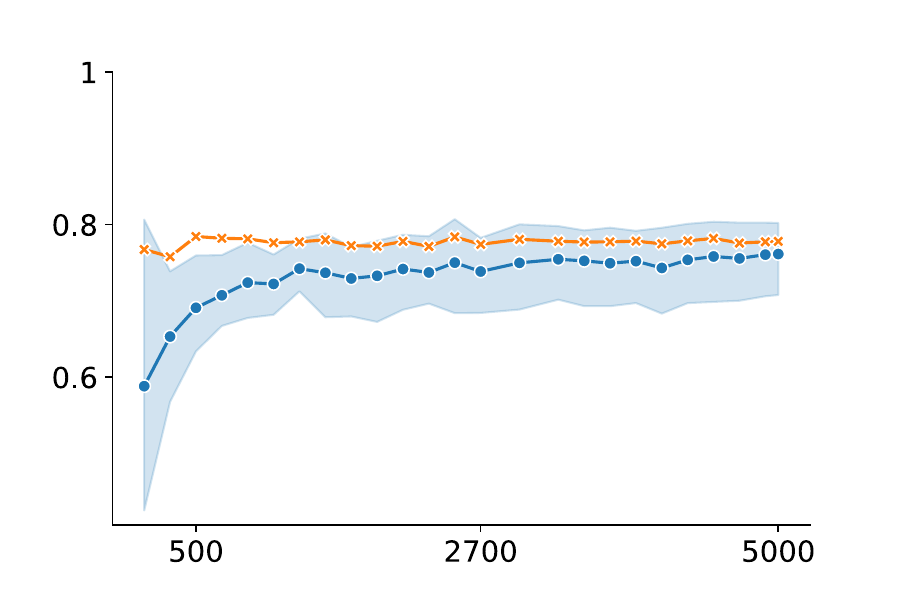}
         \caption{Linear ($\kappa=0.1$)}
     \end{subfigure}
     \begin{subfigure}{0.28\textwidth}
         \centering
         \includegraphics[width=\textwidth]{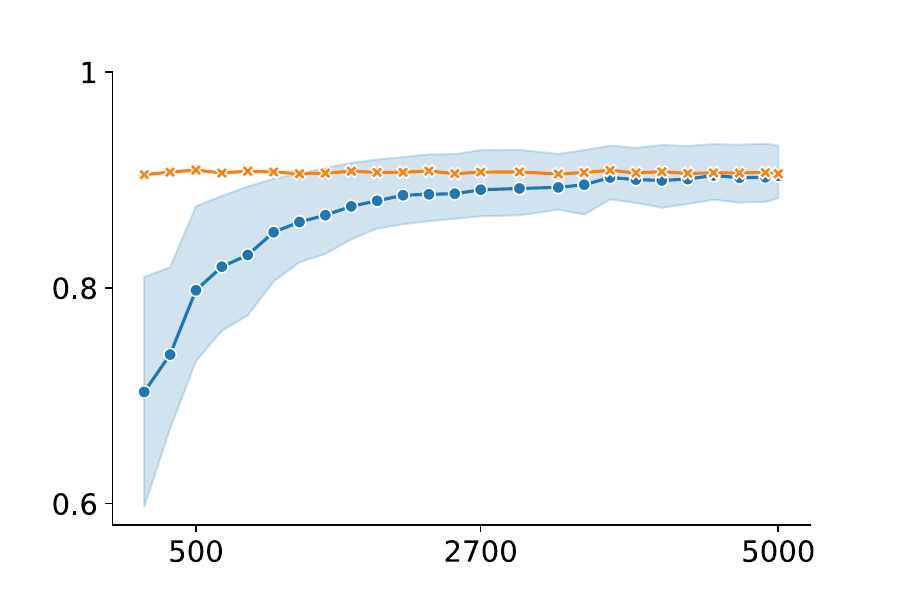}
         \caption{Exponential ($\kappa=0$)}
     \end{subfigure}
     \hspace{0.2cm}
     \begin{subfigure}{0.28\textwidth}
         \centering
         \includegraphics[width=\textwidth]{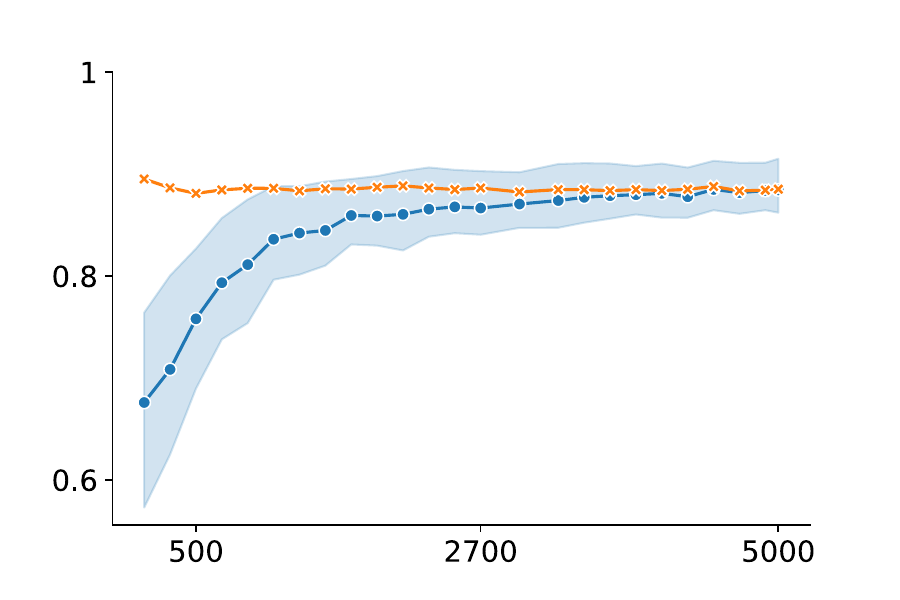}
         \caption{Exponential ($\kappa=0.05$)}
     \end{subfigure}
     \hspace{0.2cm}
     \begin{subfigure}{0.28\textwidth}
         \centering
         \includegraphics[width=\textwidth]{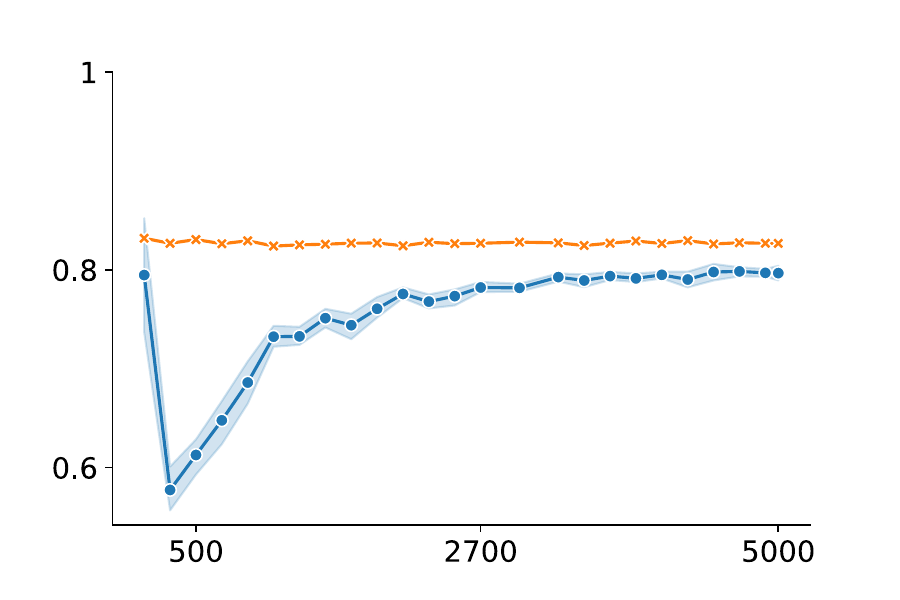}
         \caption{Exponential ($\kappa=0.1$)}
     \end{subfigure}
        \caption{Simulation results of the sample Gcorr ($\bullet $) and HSIC ($\times$). The left column denotes latent positions, $X$ and $Y$, have a linear relationship, and the right column denotes an exponential relation. The first row results from noise level 0, the second is noise level 0.05, and the third is 0.1. The vertical axis represents the  sample correlation, and the horizontal axis represents the number of graph nodes.}
        \label{fig:asym_gcorr}
\end{figure}

\noindent
\textbf{Study 1.} We use two linear relationships to verify that the sample Gcorr will converge to the HSIC directly calculated from the latent position when the number of network nodes increases. The graph's number of nodes $n$ ranges from 100 to 5000. In this study, we only consider the univariate latent positions.
Suppose the latent positions $X_{1}, \cdots,X_{n}$ are generated from a beta distribution with $\alpha=1$ and $\beta=2$, and let $\epsilon \sim \mathcal{N}(0,1)$, we consider two linear relationships of $X$ and $Y$:
\begin{itemize}
\item[(i)] \textbf{Linear}: $Y=X+\kappa\epsilon$;\quad
\item[(ii)] \textbf{Exponential}: $Y=\exp(X)+\kappa\epsilon$, 
\end{itemize}
where $\kappa$ denote the noise level.
Three noise levels are set for each relationship separately. The kernel functions for generating random graphs are Gaussian kernel $K(x,x^{'})=\mathrm{exp}\left\{-(x-x^{'})^2/\sigma ^2 \right\} $ with $\sigma=1$ and Laplace Kernel $L(y,y^{'})=\mathrm{exp}\left\{-c \left| y-y^{'} \right| \right\}$ with $c=1$. We chose the embedding dimensions $d=7$. In Figure \ref{fig:asym_gcorr}, we observe the convergence of Gcorr to the HISC computed using latent positions as the number of nodes increases. This convergence remains evident even under conditions of noise with $\kappa=0.05$ and $\kappa=0.1$, albeit potentially requiring a larger number of nodes for accurate estimation. Notably, the practical consideration that latent positions cannot be directly observed underscores the importance and validity of our proposed estimation approach utilizing graph embedding techniques. 
\graphicspath{{figs/}}

\noindent
\textbf{Study 2.} In this study, we compare the testing power and runtime for four different relationships under univariate and multivariate settings. Let $\epsilon_1=\mathcal{N}(0,0.1)$ and $\mathcal{U}(a,b)$ denote the uniform distribution on the interval $(a,b)$. For the multivariate setting, let $\mathbf{w} \in \mathbb{R}^p$ be a decaying vector where $w_i = 1/i$ for $i \in [p]$. Then we have:
\begin{itemize}
    \item[(i)] \textbf{Linear} $(X,Y)\in \mathbb{R}^p \times \mathbb{R}$: $X \sim \mathcal{U}(0,1)^p$ and $Y=\mathbf{w}^\top X + \epsilon_1$;
    \item[(ii)] \textbf{Exponential} $(X,Y)\in \mathbb{R}^p \times \mathbb{R}$: $X \sim \mathcal{U}(0,1)^p$ and $Y=\exp \left( \mathbf{w}^\top X \right) + \epsilon_1$;
    \item[(iii)] \textbf{W shape} $(X,Y)\in \mathbb{R}^p \times \mathbb{R}$: $X \sim \mathcal{U}(-1,1)^p$, $\mathbf{u} \sim \mathcal{U}(-1,1)^p$ and 
    \begin{align*}
        Y = 4 \left\{ \left[ \left( \mathbf{w}^\top X \right)^2 - \frac{1}{2} \right]^2 + \mathbf{w}^\top \mathbf{u}/500 \right\}+0.5 \epsilon_1;
    \end{align*}
    \item[(iv)] \textbf{Joint Normal} $(X,Y)\in \mathbb{R}^p \times \mathbb{R}^p$: $(X,Y) \sim \mathcal{N}(0,\Sigma)$ for $\Sigma=\left[ \left [ \mathbf{I}_p ,\rho \mathbf{J}_p \right ] ,\left[ \rho \mathbf{J}_p , \mathbf{I}_p \right]  \right]$, where $\mathbf{I}_p \in \mathbb{R}^{p \times p}$ is identity matrix, $\mathbf{J}_p \in \mathbb{R}^{p \times p}$ is the matrix of ones and $\rho=\left\{ 1,0.7,0.5 \right\}$.
\end{itemize}

\begin{figure}
		\scriptsize
		\begin{center}
			\begin{tabular}{c}
				\includegraphics[width=0.7\textwidth]{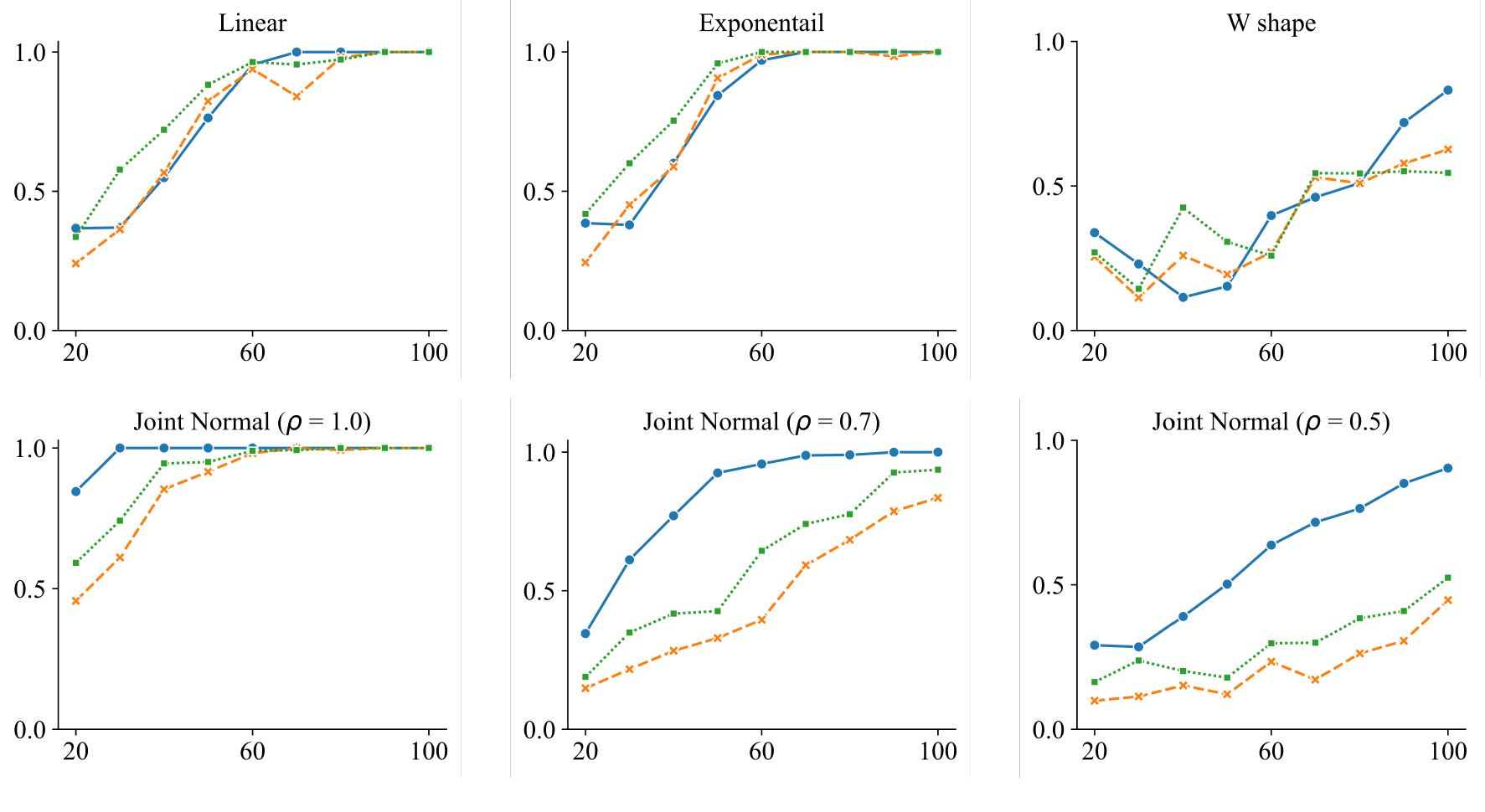}
				\\
			\end{tabular}
		\end{center}
		\caption{Comparing the testing power of Gcorr, MGC, and Dcorr for univariate simulations. The number of nodes $n$ from 20 to 100 and the Type \uppercase\expandafter{\romannumeral1} error level $\alpha=0.05$. The $\bullet$ on solid line represents Gcorr, the $\times$ on dashed line represents MGC, and the $\blacksquare$ on dotted linerepresents Dcorr. The horizontal axis represents the number of nodes, and the vertical axis represents the empirical power.}
		\label{fig:compare_uni_pow}
\end{figure}
\begin{figure}
		\scriptsize
		\begin{center}
			\begin{tabular}{c}
				\includegraphics[width=0.7\textwidth]{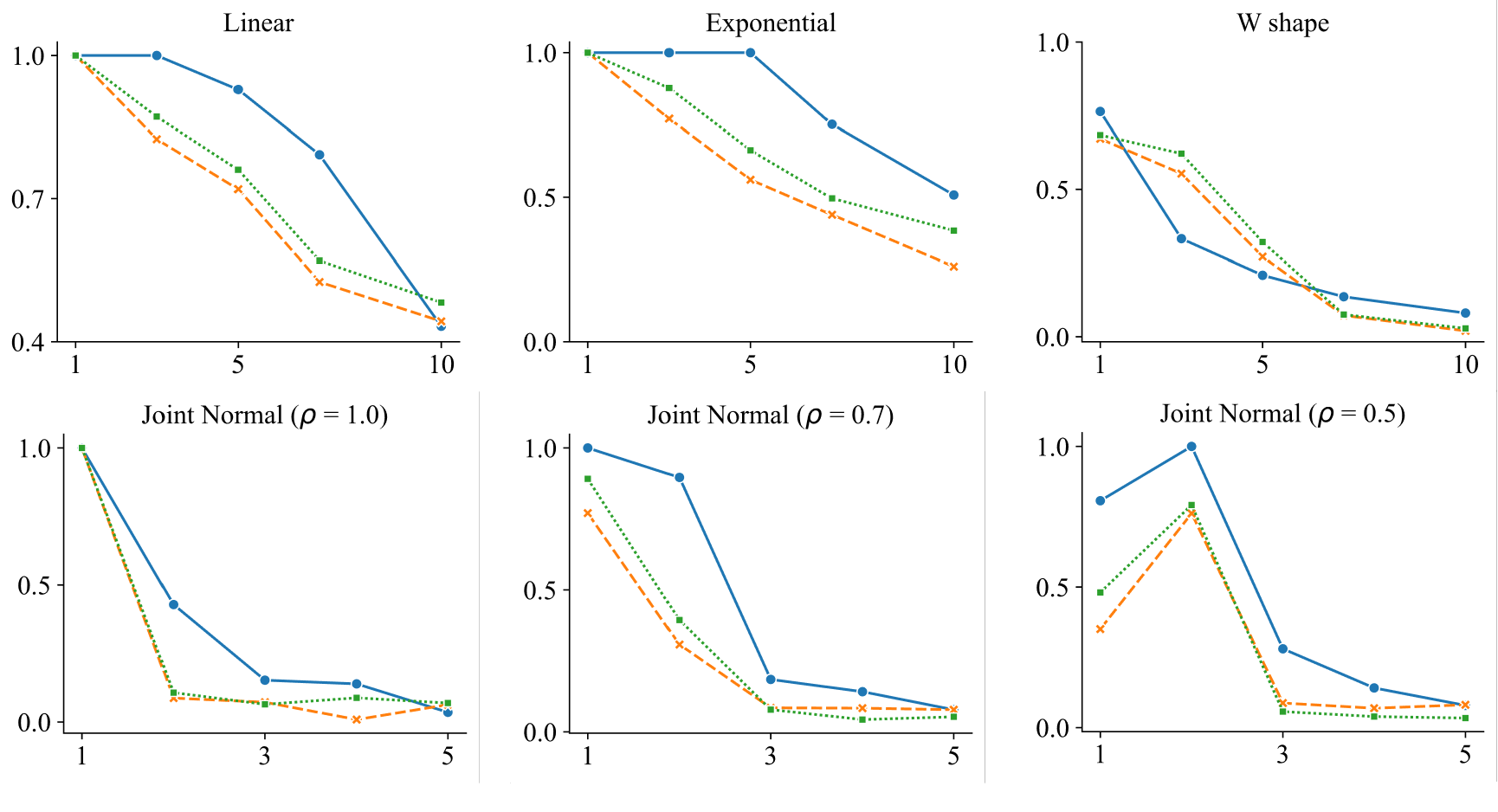}
				\\
			\end{tabular}
		\end{center}
		\caption{Comparing the testing power of Gcorr, MGC, and Dcorr for multivariate simulations. The $\bullet$ on solid line represents Gcorr, the $\times$ on dashed line represents MGC, and the $\blacksquare$ on dotted linerepresents Dcorr. The horizontal axis represents the dimension $q$, and the vertical axis represents the empirical power.}
		\label{fig:compare_multi_pow}
\end{figure}

Figure~\ref{fig:compare_uni_pow} illustrates the testing power of Gcorr and the Two-Step Correlation methods (MGC and Dcorr) for univariate latent positions under finite sample sizes. The number of nodes ranges from 20 to 100, with the testing power estimated using a permutation method. For the first two relationships, Dcorr exhibits slightly better performance than Gcorr and MGC when the number of nodes is small. However, for all three methods, the testing power under linear and exponential relationships approaches 1 as the node count increases to 100. This trend suggests that significant differences in testing power between Gcorr and Dcorr are unlikely in real-world scenarios with relatively large node counts. Notably, when the latent positions follow a joint normal distribution, Gcorr significantly outperforms the Two-Step Correlation methods (MGC and Dcorr). Figure~\ref{fig:compare_multi_pow} presents the testing power of three methods for multivariate latent positions for the number of nodes $n=100$. Gcorr consistently outperforms the other methods across all six scenarios, highlighting the advantages of the proposed approach.

\graphicspath{{figs/}}

\graphicspath{{figs/}}

\begin{figure}
		\scriptsize
		\begin{center}
			\begin{tabular}{c}
				\includegraphics[width=0.5\textwidth]{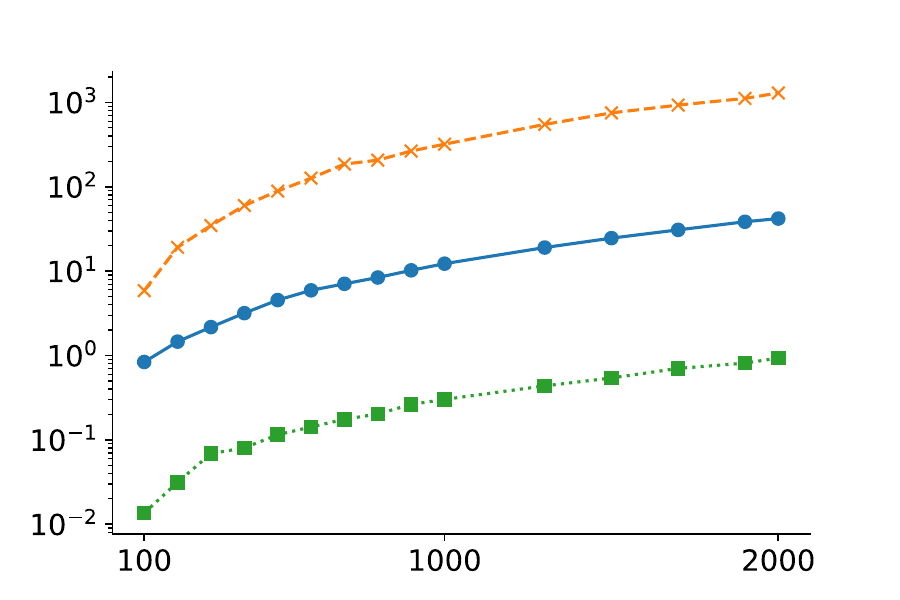}
				\\
			\end{tabular}
		\end{center}
		\caption{The running time of Gcorr, MGC, and Dcorr for different nodes. The node number   $n$ is from 100 to 2000. The $\bullet$ represents Gcorr, the $\times$ represents MGC, and the $\blacksquare$ represents Dcorr. The horizontal and vertical  axes  represent respectively the number of nodes and  the logscale running time.}
		\label{fig:compare_time}
\end{figure}

Figure~\ref{fig:compare_time} shows the runtime of calculating the p-value of Gcorr, Two-Step Correlation (MGC and Dcorr). The p-value of MGC is calculated using the \textit{multiscale\_graphcorr} function of the \textit{scipy.stats} package in Python, the p-value of Dcorr is calculated using the \textit{dcor} package in Python and the timing uses the \textit{time} package. Both MGC and Dcorr rely on a two-step estimation approach, involving initial eigendecomposition of the normalized Laplacian of the random graph to obtain low-dimensional vector representations, followed by independence testing between these vectors. Of the three methods considered, Dcorr is observed to be the fastest, with MGC being the slowest and Gcorr exhibiting intermediate speed relative to MGC.

Overall, the numerical simulations indicate that $\mathrm{gCor}$ is particularly effective when graph dependence arises from nonlinear generating mechanisms, including settings with jointly Gaussian latent positions where the induced dependence for graph level is inherently nonlinear. In contrast, alternative methods may perform comparably only in very small, low-dimensional settings with approximately linear dependence, with their advantage diminishing as graph size or dependence complexity increases.

\graphicspath{{figs/}}

\subsection{Real Data Analysis}

In this section, we use two cases to show the application of our proposed method to the real data analysis.

\subsubsection{3D mesh data analysis}

We demonstrate the applicability of our method to 3D shape comparison using the animal posture dataset of \citet{sumner2004deformation}. Mesh surfaces are constructed from the 3D point clouds via k-nearest-neighbor graphs, yielding k-regular graph representations suitable for our graph correlation analysis. For each 3D subject, we sampled $n=1000$ points using a fixed set of vertex indices, leveraging the anatomical correspondence of dataset for pairs of same class. To establish a rigorous baseline for comparisons between different class, 3D objects of different species were explicitly aligned using an optimal transport–based matching approach implemented via the Sinkhorn algorithm \citep{cuturi2013sinkhorn}. We selected $k=20$ nearest neighbors to construct the k-NN graph. Because the graph size is large ($n = 1000$), the MGC method is computationally expensive. We therefore compare our proposed $\mathrm{Gcorr}$ with $\mathrm{Dcorr}$ in this section, using an embedding dimension of $d = 10$. 
\graphicspath{{figs/}}
	\begin{figure}[h!]
		\scriptsize
		\begin{center}
			\begin{tabular}{c}
				\psfig{figure=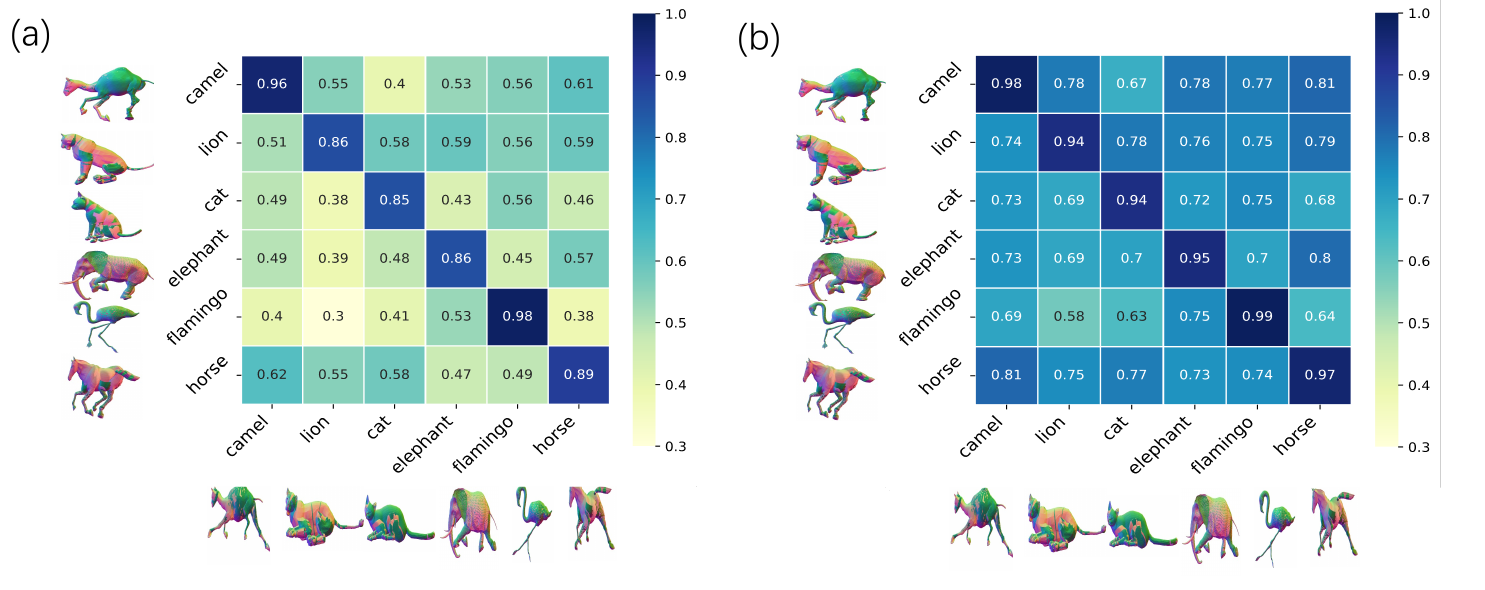,width=5in,angle=0} \\
                (A): Heatmap of normal posture \\
                \psfig{figure=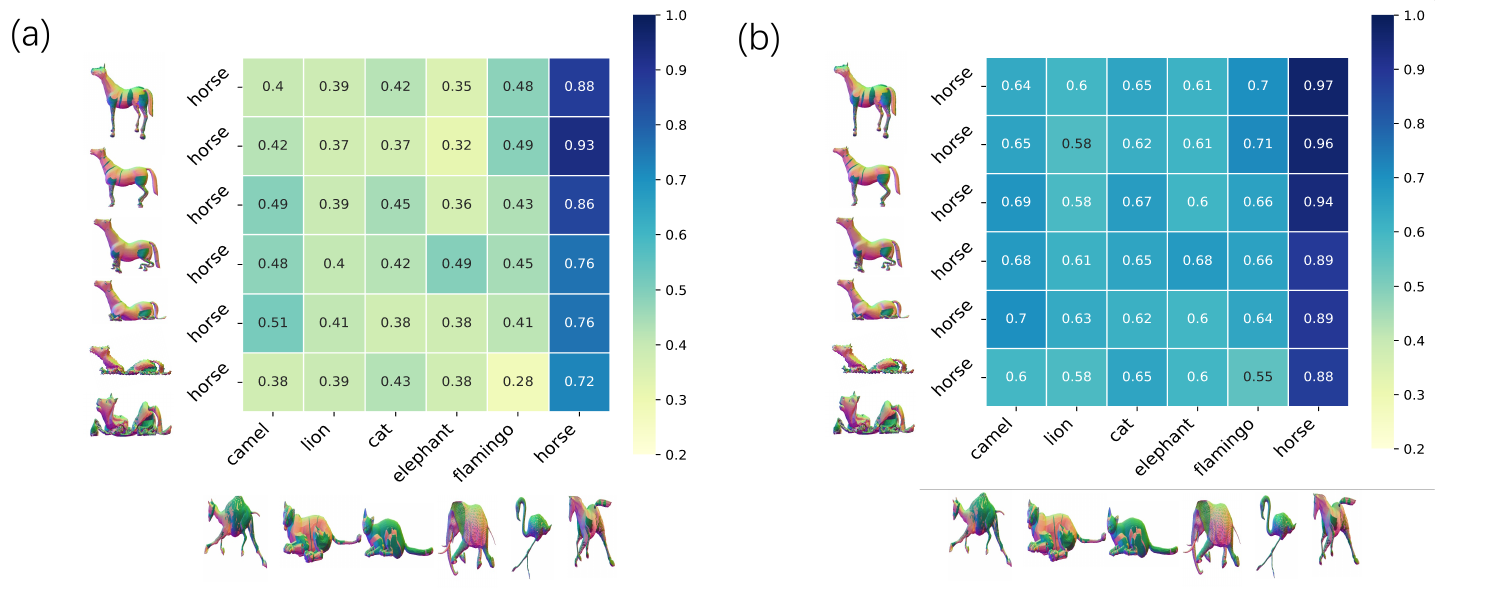,width=5in,angle=0} \\
                (B): Heatmap of collapse posture\\
			\end{tabular}
		\end{center}
		\caption{Correlation heatmaps comparing different postures across animals. For each subfigure, panel (a) shows the results based on $\mathrm{Gcorr}$, and panel (b) shows the results based on $\mathrm{Dcorr}$.} 
		\label{fig:compare_mesh_data}
	\end{figure}
Fig.~\ref{fig:compare_mesh_data} (A) shows a comparison of normal postures for six animal species. Our method effectively captures 3D shape similarities, as evidenced by the high correlation coefficients between different postures of the same species. Although $\mathrm{Dcorr}$ yields relatively large correlations for the same animal, $\mathrm{Gcorr}$ produces substantially smaller correlations across different animals. This indicates that the graph-based correlation more effectively discriminates among distinct 3D subjects, rather than uniformly assigning high correlation values across all shapes. This finding is further supported by Figure~\ref{fig:compare_mesh_data}(B), even for collapsed horse postures that are nearly indistinguishable to the human eye, our method maintains high correlations within same species while preserving low correlations across different species.

\subsubsection{Neuronal connectivity network analysis}

\begin{figure}
		\scriptsize
		\begin{center}
			\begin{tabular}{c}
				\psfig{figure=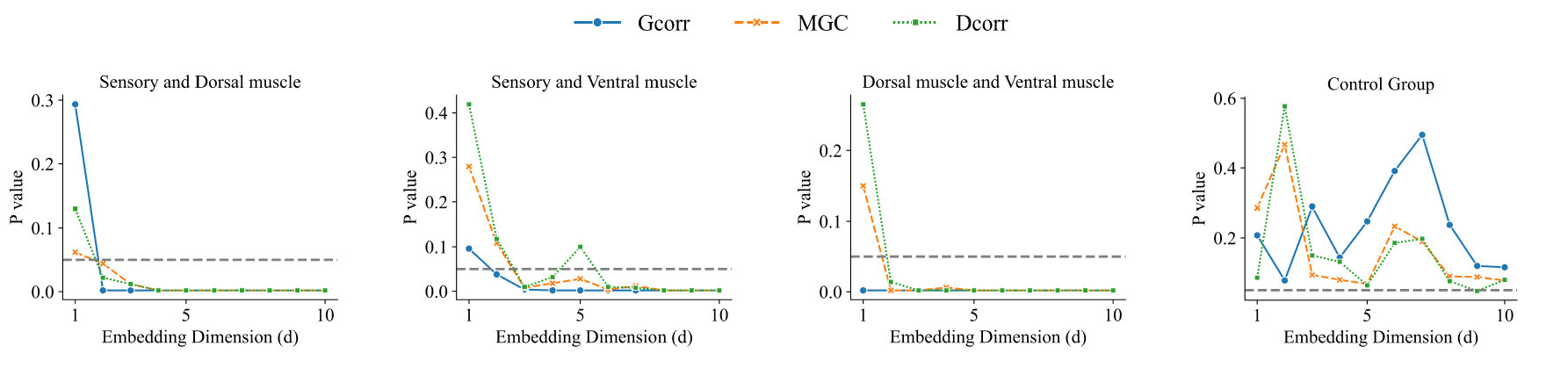,width=6.5in,angle=0} \\
			\end{tabular}
		\end{center}
		\caption{Comparison of graph correlation coefficients and p-values. The permutation test was repeated $m=500$. The $\bullet$ represents Gcorr, the $\times$ represents MGC, and the $\blacksquare$ represents Dcorr. The horizontal and vertical  axes  represent respectively the embedding dimension and the p-value of 0.05.} 
		\label{fig:real_data_pval}
	\end{figure}

The neuronal connectivity network of the hermaphrodite C. elegans was used as the dataset in this study, previously compiled and organized by \cite{chen2006wiring,varshney2011structural}. This network comprises 200 sensory and motor neurons that connect 20 sensory features and 95 body wall muscles. The complete neuronal network can be categorized into three sub-networks: sensory control, dorsal body wall muscles, and ventral body wall muscles. To assess network independence among the three sub-networks, we computed graph correlations between pairs of sub-networks: sensory-dorsal body wall muscle, sensory-ventral body wall muscle, and dorsal-ventral body wall muscle. The three subnetworks are matched based on adjacency spectral embedding, which is well suited for latent position random graph models and does not rely on geometric information \citep{lyzinski2015graph}. We employed a permutation test to evaluate statistical significance, varying the embedding dimension $d$ from 1 to 10. As a control, we compared the sensory sub-network to a random Erd\H{o}s-R\'{e}nyi graph. We compare the p-values obtained by our method with those from $\mathrm{MGC}$ and $\mathrm{Dcorr}$. As shown in Figure~\ref{fig:real_data_pval}, all three approaches detect strong dependence among the three subnetworks, with p-values consistently below 0.05 across embedding dimensions. Notably, our method exhibits more robust independence testing performance at smaller embedding dimensions.
Intriguingly, the sensory subnetwork exhibits a stronger correlation with the dorsal muscle subnetwork than the ventral muscle subnetwork. This disparity likely stems from the dorsal muscles' greater involvement in sensory processing and response \citep{schafer2014mechanosensory, goodman2019caenorhabditis} compared to the ventral muscles, which are more specialized for rhythmic locomotion \citep{gjorgjieva2014neurobiology}. These findings underscore the functional specialization of the dorsal and ventral muscle sub-networks.


\graphicspath{{figs/}}


\section{Discussion}\label{sec6}

This paper formalizes the population-level correlation of latent position random graphs and provides a consistent sample-based estimator via spectral decomposition of the adjacency matrix. We establish convergence of the sample estimator to its population counterpart and derive the asymptotic distribution of the sample graph covariance under suitable conditions. The proposed $\mathrm{gCor}$ method demonstrates strong empirical performance while maintaining favorable computational complexity. Our theoretical guarantees rely on the assumption that the graph-generating kernels are universal. If the generating kernel is not universal or cannot be well approximated by a universal kernel, the method remains well defined but universal consistency is no longer guaranteed. This highlights an important practical consideration when applying the method to graphs with unknown or restricted generating mechanisms. Future research directions include extending the framework to more general inhomogeneous random graph models via estimation of scaling parameters, improving testing power in small-node regimes and multivariate latent positions, and developing principled strategies for selecting the embedding dimension. These extensions would further broaden the applicability and robustness of random graph dependence testing.

\section{Disclosure statement}\label{disclosure-statement}

The authors report there are no competing interests to declare.

\newpage

\appendix

\bigskip
\begin{center}
{\large\bf SUPPLEMENTARY MATERIAL}
\end{center}

The supplementary material are divided into three parts, the Section~\ref{sec:external_result} contains some auxiliary lemmas we will use in the later proof, Section~\ref{sec:internal_result} includes all the proofs for theorems and technical lemmas and Section~\ref{sec:additional_sim} contains additional simulation results.

\section{External Results}\label{sec:external_result}

\begin{lemma}[Result \uppercase\expandafter{\romannumeral1} in \cite{tosio1973continuity}]\label{bound_psd_trans}

For any two bounded linear operator $A,B:\mathcal{H}_1 \to \mathcal{H}_2$, where $\mathcal{H}_1$ and $\mathcal{H}_2$ are two Hilbert spaces, let $\left| M \right| \defn (M^* M)^{1/2}$ where $M \in \{ A, B\}$, the map $\left| \cdot \right|$ is almost Lipschitz-continuous, such that
\begin{align*}
    \left\|  \left| A \right| - \left| B \right|  \right\| \le \frac{2}{\pi} \left[ 2 + \log \left( \frac{\left| A \right|+ \left| B \right|}{\left\|   A  - B  \right\|} \right) \right] \left\|  A  - B  \right\|,
\end{align*}
where $\left\| \cdot  \right\|$ denotes the spectral norm.
   
\end{lemma}

\begin{lemma}[Theorem 4.1 in \cite{sterge2020gain}]\label{bound_kpca}

Suppose $\mathcal{X}$ is a separable topological space and $\mathcal{H}$ is a separable RKHS with bounded, continuous and positive definite kernel $K$. Let the (uncentered) covariance operator on $\mathcal{H}$ defined as $C \defn \int_{\mathcal{X}} K(\cdot,x) \otimes K(\cdot,x) \mathrm{d} \mathbb{P}(x)$. Assume the eigenvalues $(\lambda_i)_{i \in I}$ of $C$ are simple, positive and arranged as a non-increasing way, i.e. $\lambda_1 > \lambda_2 > \cdots$. To project the kernel $K$ onto $d$ dimensional subspace, the reconstruction error is defined as
\begin{align*}
    R(P_d) = \left\| (I-P_d)C^{1/2} \right\|^2_{\mathrm{HS}(\mathcal{H},\mathcal{H})}.
\end{align*}
For any $t>0$, defined the efficient dimension as $\mathcal{N}_C(t) = \mathrm{tr} \left[ (C+tI)^{-1}C  \right]$. Suppose $n >3$, $0<\delta <1$, $t = \Theta \left[\frac{9}{n} \log (n / \delta) \right] $, then we have
\begin{align}\label{eq:bound_kernel_pca} 
\mathbb{P}^n \left\{ (X_i)_{i=1}^n: R(P_d) \le 9 \mathcal{N}_C(t)(\lambda_d + t)   \right\} \ge 1-\delta.
\end{align}
    
\end{lemma}

\begin{lemma}[Lemma C.9 in \cite{sriperumbudur2022approximate}]\label{order_Nc}

Let $\mathcal{X}$, $\mathcal{H}$, $K$, $C$, $(\lambda_i)_{i \in I}$ and $\mathcal{N}_C(t)$ defined as in Lemma~\ref{bound_kpca}, the following results hold:
\begin{itemize}
    \item[(i)] If $\lambda_i =\mathcal{O} (i^{-\alpha})$ for $\alpha >1$, then $\mathcal{N}_{C}(t) := \mathrm{tr} \left(C(C+tI)^{-1}\right) = \mathcal{O} \left(t^{-\frac{1}{\alpha}} \right)$, where $t>0$.
    \item[(ii)] If $\lambda_i=\mathcal{O}(e^{-\tau i})$ for some $\tau >0$, then $\mathcal{N}_{C}(t) = \mathcal{O}\left( \log \left( \frac{1}{t} \right)\right)$, where $t>0$.
\end{itemize}
    
\end{lemma}

\begin{lemma}[Theorem 3.1 in \cite{tang2013universally}]\label{concentration_adj}

Suppose the LPRG $G=(V,E)$ generated by the latent positions $\X \in \mathbb{R}^{n \times q} \subset \mathcal{X}$ and the kernel function $K:\mathcal{X} \times \mathcal{X} \to [0,1]$, giving rise to the Gram matrix $\mathbf{K} = \left[ K_{i,j} \right]_{i,j=1}^n = \left[ K(\x_i,\x_j) \right]_{i,j=1}^n \in \mathbb{R}^{n \times n}$. The associated adjacency matrix $\A = \left[ A_{i,j} \right]_{i,j=1}^n \in \mathbb{R}^{n \times n}$ is a symmetric matrix with its upper diagonal elements follows a conditionally independent Bernoulli distribution such that $A_{i,j} \sim \mathrm{Bernoulli}(K_{i,j})$ for $i<j$ and $A_{i,i}=0$ for $i = 1,\dots,n$. Let the embedding dimension $d \ge 1$. Let the eigenvalues of the integral operator $T_K$ arranged as a non-increasing way and denoted as $\lambda_1 \ge \lambda_2 \ge \cdots$, the $d$-th eigen gap of $T_K$ is denoted as $\delta_d \defn \lambda_d - \lambda_{d+1}$ ans we assume $\delta_d \ge 0$. Let the spectral decomposition of $\A$ represented as $\A = \U \bS \U^\top \approx \U_d \bS_d \U_d^\top$, where $\bS_d \in \mathbb{R}^{d \times d}$ composed by the first $d$ eigenvalues of $\A$ and $\U_d \in \mathbb{R}^{n \times d}$ are associated eigenvectors. Similarly, for $\mathbf{K}$ we have $\mathbf{K} = \U_{K} \bS_{K} \U_{K}^\top \approx \U_{K,d} \bS_{K,d} \U_{K,d}^\top$, where $\bS_{K,d} \in \mathbb{R}^{d \times d}$ composed by the first $d$ eigenvalues of $\mathbf{K}$ and $\U_{K,d} \in \mathbb{R}^{n \times d}$ are associated eigenvectors. Let $\Phi_d \defn \U_{K,d} \bS_{K,d}^{1/2} \in \mathbb{R}^{n \times d}$, then there exists a unitary matrix $\W \in \mathbb{R}^{d \times d}$ and $\hat{\Phi}_d \defn \U_d \bS_d^{1/2} \W \in \mathbb{R}^{n \times d}$ such that for any $\varepsilon > 0$
\begin{align*}
    \mathbb{P} \left(  \left \| \hat{\phi}_d(x_i) - \phi_d(x_i) \right \|_2 \ge \varepsilon   \right) \le 27 \delta_d^{-2}\varepsilon^{-1} \sqrt{6d \log (n)/n} ,
\end{align*}
where $\hat{\phi}_d(x_i)$ and $\phi_d(x_i)$ denote the $i$-th row of $\hat{\Phi}_d$ and $\Phi_d$, respectively. Besides, we also have
\begin{align}\label{bound_adj_appro}
    \mathrm{E} \left[ \left \| \hat{\phi}_d(x_i) - \phi_d(x_i) \right \|_2  \right] \le 27 \delta_d^{-2} \sqrt{6d \log (n)/n} .
\end{align}
   
\end{lemma}

\begin{remark}
    Equation \eqref{bound_adj_appro} in Lemma~\ref{concentration_adj} is not stated in Theorem 3.1 but rather appears as equation (3.9) within the proof of Theorem 3.1.
\end{remark}

\begin{lemma}[Lemma 6 in \cite{levin2021limit}]\label{concentration_laplace}

Suppose the LPRG $G$, $\X$, $\mathbf{K}$ and $\A$ defined as in Lemma~\ref{concentration_adj}. Let the embedding dimension $d \ge 1$. Let the normalized Laplacian matrices of $\A$ and $\mathbf{K}$ denoted as $\mathbf{L}(\A) \in \mathbb{R}^{n \times n}$ and $\mathbf{L}(\mathbf{K}) \in \mathbb{R}^{n \times n}$, respectively. The $d$-th truncated spectral decomposition of $\mathbf{L}(\A)$ can be represented as $\mathbf{L}(\A) \approx \tilde{\U}_{A,d} \tilde{\bS}_{A,d} \tilde{\U}_{A,d}^\top$, where $\tilde{\bS}_{A,d} \in \mathbb{R}^{d \times d}$ composed by the first $d$ eigenvalues of $\mathbf{L}(\A)$ and $\tilde{\U}_{A,d} \in \mathbb{R}^{n \times d}$ are associated eigenvectors. Similarly, we have $\mathbf{L}(\mathbf{K}) \approx \tilde{\U}_{K,d} \tilde{\bS}_{K,d} \tilde{\U}_{K,d}^\top$, where $\tilde{\bS}_{K,d} \in \mathbb{R}^{d \times d}$ composed by the first $d$ eigenvalues of $\mathbf{L}(\mathbf{K})$ and $\tilde{\U}_{K,d} \in \mathbb{R}^{n \times d}$ are associated eigenvectors. Then there exists an orthogonal matrix $\mathbf{Q} \in \mathbb{R}^{d \times d}$ such that
\begin{align*}
    \left \| \tilde{\U}_{A,d} \tilde{\bS}_{A,d}^{1/2} - \tilde{\U}_{K,d} \tilde{\bS}_{K,d}^{1/2} \mathbf{Q} \right \|_{2,\infty} = \mathcal{O} \left( \log^{1/2}(n) /n\right) ,
\end{align*}
where $\left \| \cdot \right \|_{2,\infty}$ denotes the 2-to-$\infty$ norm such that $\left \| M \right \|_{2,\infty} = \sup_{x: \left\| x \right\|=1} \left\| Mx \right\|_{\infty}$.
   
\end{lemma}

\section{Internal Results}\label{sec:internal_result}

\begin{proof}[Proof of Theorem 1]

\textbf{Step 1: Definition of truncated graph covariance.} First we introduce the Mercer's representation theorem in RKHS, and define the truncated population graph covariance. From our previous definition, the population graph covariance $\mathrm{hCov}(G_{1},G_{2})$ of two random graphs can be written as
\begin{align}\label{pop_hcov}
     \mathrm{hCov}(G_{1},G_{2}) & = \mathrm{E}\left [ K(X,{X}')L(Y,{Y}') \right ]+\mathrm{E}\left [ K(X,{X}') \right ]\mathrm{E}\left [ L(Y,{Y}') \right ] \nonumber \\
     & \quad -2\mathrm{E}\left [ K(X,{X}')L(Y,{Y}'') \right ].
\end{align}
Assume we have a compact metric space $(\mathcal{X},d)$ and $\nu$ denotes the strictly positive and finite Borel $\sigma$-field of $\mathcal{X}$. In the calculation equation of $\mathrm{hCov}(G_{1},G_{2})$, the $K:\mathcal{X} \times \mathcal{X}\rightarrow [0,1]$ is continuous and positive definite kernel function. 
Then define the space of square-integrable functions on $\nu$ as $L^2(\mathcal{X},\nu)$, the integral operator $T_{K}:L^2(\mathcal{X},\nu) \rightarrow L^2(\mathcal{X},\nu)$ defined by:
\begin{equation*}
    (T_{K}f)(x) = \int _{\mathcal{X}}K(x,{x}')f({x}')d\nu ({x}'), \quad f\in L^2(\mathcal{X},\nu).
\end{equation*}
$T_{K}$ is a continuous and compact operator. Let $\left \{ \lambda _{i} \right \}$ be the set of eigenvalues of $T_K$ and $\lambda _{1} \geq \lambda _{2} \geq \cdots \geq 0$. $\left \{ \psi _{i} \right \}$ be the set of corresponding eigenfunctions.

Considering the feature mapping of a kernel which takes the form 
\begin{equation*}
    \left \langle \phi(x),\phi({x}')  \right \rangle_{\mathcal{H}_K}=K(x,{x}'),
\end{equation*}
the Mercer's representation theorem shows that the feature map $\phi(x)$ can be denoted as
\begin{equation*}
    \phi(x) = (\sqrt{\lambda_{j}}\psi _{j}(x):j=1,2,\cdots).
\end{equation*}
Let $\phi_{d}(x)$ as the truncation of $\phi(x)$ to $\mathbb{R}^d$, where
\begin{equation*}
    \phi_{d}(x) = (\sqrt{\lambda_{j}}\psi _{j}(x):j=1,2,\cdots,d).
\end{equation*}
Let $K_{d}(x,{x}')=\left \langle \phi_{d}(x),\phi_{d}({x}')  \right \rangle_{\mathcal{H}_K} \subset \mathcal{H}_K$ represents the d-th truncation of the kernel space. Then there exists an orthogonal projection $P_d$ such that $K_d (\cdot, x) = P_d K(\cdot,x)$ for all $x \in \mathcal{X}$.

The d-th truncation of $\mathrm{hCov}(G_{1},G_{2})$ is $\mathrm{hCov}^{(d)}(G_{1},G_{2})$, then the truncated population graph covariance can be defined as follows:
\begin{align}\label{pop_truncated_hcov}
     \mathrm{hCov}^{(d)}(G_{1},G_{2}) &= \mathrm{E}\left [ K_{d}(X,{X}')L_{d}(Y,{Y}') \right ]+\mathrm{E}\left [ K_{d}(X,{X}') \right ]\mathrm{E}\left [ L_{d}(Y,{Y}') \right ] \nonumber\\
     & -2\mathrm{E}\left [ K_{d}(X,{X}')L_{d}(Y,{Y}'') \right ].
\end{align}

\textbf{Step 2. Consistency of truncated graph covariance.} In this step, we want to show the $\mathrm{hCov}^{(d)}(G_{1},G_{2})$ will converge to $\mathrm{hCov}(G_{1},G_{2})$ under the specific range of embedding dimension $d$. 

We can write \eqref{pop_hcov} as the form of kernel mean embedding such as
\begin{align*}
    \mathrm{hCov}(G_1,G_2) = \left\| \mu_G(\mathbb{P}_{XY}) -  \mu_G(\mathbb{P}_{X} \otimes \mathbb{P}_{Y}) \right\|^2_{\mathcal{H}},
\end{align*}
where $\mathcal{H} = \mathcal{H}_K \otimes \mathcal{H}_L$ denote the tensor product Hilbert space associated with the tensor product kernel $G = K \otimes L$, the kernel mean embedding is defined as $\mu_G(\mathbb{P}) = \int_{\mathcal{X}} G(\cdot,x) \mathrm{d} \mathbb{P} (x)$ for all $x \in \mathcal{X}$. Similarly, for the $\mathrm{hCov}^{(d)}(G_1,G_2)$, we can define the truncated tensor product kernel $G_d = K_d \otimes L_d$ and the associated tensor product Hilbert space $\mathcal{H} = \mathcal{H}_{K} \otimes \mathcal{H}_{L}$, and thus we got
\begin{align*}
    \mathrm{hCov}^{(d)}(G_1,G_2) = \left\| \mu_{G_d}(\mathbb{P}_{XY}) -  \mu_{G_d}(\mathbb{P}_{X} \otimes \mathbb{P}_{Y}) \right\|^2_{\mathcal{H}}.
\end{align*}

Then we have
\begin{align*}
    & \quad  \mathrm{hCov}(G_1,G_2) - \mathrm{hCov}^{(d)}(G_1,G_2) \\
    & =  \left\| \mu_G(\mathbb{P}_{XY}) -  \mu_G(\mathbb{P}_{X} \otimes \mathbb{P}_{Y}) \right\|^2_{\mathcal{H}} - \left\| \mu_{G_d}(\mathbb{P}_{XY}) -  \mu_{G_d}(\mathbb{P}_{X} \otimes \mathbb{P}_{Y}) \right\|^2_{\mathcal{H}} \\
    & \stackrel{(a)}{=} \left \langle \mu_G(\mathbb{P}_{XY}) -  \mu_G(\mathbb{P}_{X} \otimes \mathbb{P}_{Y}), \mu_G(\mathbb{P}_{XY}) -  \mu_G(\mathbb{P}_{X} \otimes \mathbb{P}_{Y}) \right \rangle_{\mathcal{H}} \\
    & \quad - \left \langle \mu_{G_d}(\mathbb{P}_{XY}) -  \mu_{G_d}(\mathbb{P}_{X} \otimes \mathbb{P}_{Y}) , \mu_{G_d}(\mathbb{P}_{XY}) -  \mu_{G_d}(\mathbb{P}_{X} \otimes \mathbb{P}_{Y}) \right \rangle_{\mathcal{H}} \\
    & \stackrel{(b)}{=} \left \langle \mu_G(\mathbb{P}_{XY}) -  \mu_G(\mathbb{P}_{X} \otimes \mathbb{P}_{Y}), \mu_G(\mathbb{P}_{XY}) -  \mu_G(\mathbb{P}_{X} \otimes \mathbb{P}_{Y}) \right \rangle_{\mathcal{H}} \\
    & \quad - \left \langle \mu_G(\mathbb{P}_{XY}) -  \mu_G(\mathbb{P}_{X} \otimes \mathbb{P}_{Y}), \mu_{G_d}(\mathbb{P}_{XY}) -  \mu_{G_d}(\mathbb{P}_{X} \otimes \mathbb{P}_{Y}) \right \rangle_{\mathcal{H}} \\
    & \quad + \left \langle \mu_G(\mathbb{P}_{XY}) -  \mu_G(\mathbb{P}_{X} \otimes \mathbb{P}_{Y}), \mu_{G_d}(\mathbb{P}_{XY}) -  \mu_{G_d}(\mathbb{P}_{X} \otimes \mathbb{P}_{Y}) \right \rangle_{\mathcal{H}} \\
    & \quad - \left \langle \mu_{G_d}(\mathbb{P}_{XY}) -  \mu_{G_d}(\mathbb{P}_{X} \otimes \mathbb{P}_{Y}) , \mu_{G_d}(\mathbb{P}_{XY}) -  \mu_{G_d}(\mathbb{P}_{X} \otimes \mathbb{P}_{Y}) \right \rangle_{\mathcal{H}} \\
    & \stackrel{(c)}{=} \left \langle \mu_G(\mathbb{P}_{XY}) -  \mu_G(\mathbb{P}_{X} \otimes \mathbb{P}_{Y}), \left[ \mu_G(\mathbb{P}_{XY}) -  \mu_G(\mathbb{P}_{X} \otimes \mathbb{P}_{Y}) \right] \right.\\
    & \left. \quad - \left[ \mu_{G_d}(\mathbb{P}_{XY}) -  \mu_{G_d}(\mathbb{P}_{X} \otimes \mathbb{P}_{Y})  \right] \right \rangle_{\mathcal{H}} \\
    & \quad + \left \langle \left[ \mu_G(\mathbb{P}_{XY}) -  \mu_G(\mathbb{P}_{X} \otimes \mathbb{P}_{Y}) \right] - \left[ \mu_{G_d}(\mathbb{P}_{XY}) -  \mu_{G_d}(\mathbb{P}_{X} \otimes \mathbb{P}_{Y})  \right], \right.\\
    & \left. \quad \quad \mu_{G_d}(\mathbb{P}_{XY}) -  \mu_{G_d}(\mathbb{P}_{X} \otimes \mathbb{P}_{Y}) \right \rangle_{\mathcal{H}} \\
    & \stackrel{(d)}{\le} \left\| \mu_G(\mathbb{P}_{XY}) -  \mu_G(\mathbb{P}_{X} \otimes \mathbb{P}_{Y})  \right\|_{\mathcal{H}} \underbrace{\left\| \mu_G(\mathbb{P}_{XY}) - \mu_{G_d}(\mathbb{P}_{XY})  - \left[ \mu_G(\mathbb{P}_{X} \otimes \mathbb{P}_{Y}) - \mu_{G_d}(\mathbb{P}_{X} \otimes \mathbb{P}_{Y}) \right]  \right\|_{\mathcal{H}}}_{=:A_1} \\
    & \quad + \underbrace{\left\| \mu_G(\mathbb{P}_{XY}) - \mu_{G_d}(\mathbb{P}_{XY})  - \left[ \mu_G(\mathbb{P}_{X} \otimes \mathbb{P}_{Y}) - \mu_{G_d}(\mathbb{P}_{X} \otimes \mathbb{P}_{Y}) \right]  \right\|_{\mathcal{H}}}_{=:A_1} \\
    & \quad \quad \times \left\| \mu_{G_d}(\mathbb{P}_{XY}) -  \mu_{G_d}(\mathbb{P}_{X} \otimes \mathbb{P}_{Y})  \right\|_{\mathcal{H}}. 
\end{align*}
(a) is followed from the definition of square of norm, (b) is obtained by adding the item $\left \langle \mu_G(\mathbb{P}_{XY}) -  \mu_G(\mathbb{P}_{X} \otimes \mathbb{P}_{Y}), \mu_{G_d}(\mathbb{P}_{XY}) -  \mu_{G_d}(\mathbb{P}_{X} \otimes \mathbb{P}_{Y}) \right \rangle_{\mathcal{H}}$, (c) is implied by combining some similar items and (d) is because $\left \langle A,B \right \rangle_{\mathcal{H}} \le \left \| A \right \|_{\mathcal{H}} \left \| B \right \|_{\mathcal{H}} $. Once we can prove the term $A_1$ goes to 0 as $n \to \infty$, we can say that $\mathrm{hCov}^{(d)}(G_{1},G_{2})$ will converge to $\mathrm{hCov}(G_{1},G_{2})$. 

For $A_1$ we have
\begin{align*}
    A_1 \stackrel{(a)}{\le} \underbrace{\left\| \mu_{G}(\mathbb{P}_{XY})-\mu_{G_d}(\mathbb{P}_{XY}) \right\|_{\mathcal{H}}}_{=:B_1} + \underbrace{\left\| \mu_G(\mathbb{P}_{X} \otimes \mathbb{P}_{Y}) - \mu_{G_d}(\mathbb{P}_{X} \otimes \mathbb{P}_{Y}) \right\|_{\mathcal{H}}}_{=:B_2}
\end{align*}
(a) follows from the triangular inequality. Then we analysis the convergence property of $B_1$ and $B_2$ separately. We start from $B_2$:
\begin{itemize}
    \item \textbf{Term $B_2$}: For the $B_2$, we have
    \begin{align*}
        \mu_G (\mathbb{P}_X \otimes \mathbb{P}_Y) = \mu_K (\mathbb{P}_X) \otimes \mu_L (\mathbb{P}_Y), \quad \mu_{G_d} (\mathbb{P}_X \otimes \mathbb{P}_Y) = \mu_{K_d} (\mathbb{P}_X) \otimes \mu_{L_d} (\mathbb{P}_Y).
    \end{align*}
    Thus we got
    \begin{align*}
        B_2 & = \left\| \mu_K (\mathbb{P}_X) \otimes \mu_L (\mathbb{P}_Y) - \mu_{K_d} (\mathbb{P}_X) \otimes \mu_{L_d} (\mathbb{P}_Y)  \right\|_{\mathcal{H}} \\
        & \stackrel{(a)}{=} \left\| \mu_K (\mathbb{P}_X) \otimes \mu_L (\mathbb{P}_Y) - \mu_{K_d} (\mathbb{P}_X) \otimes \mu_L (\mathbb{P}_Y) + \mu_{K_d} (\mathbb{P}_X) \otimes \mu_L (\mathbb{P}_Y) \right. \\
        & \quad \left. - \mu_{K_d} (\mathbb{P}_X) \otimes \mu_{L_d} (\mathbb{P}_Y)  \right\|_{\mathcal{H}} \\
        & \stackrel{(b)}{\le } \left\| \mu_K (\mathbb{P}_X) \otimes \mu_L (\mathbb{P}_Y) - \mu_{K_d} (\mathbb{P}_X) \otimes \mu_{L} (\mathbb{P}_Y)  \right\|_{\mathcal{H}} \\
        & \quad + \left\| \mu_{K_d} (\mathbb{P}_X) \otimes \mu_{L} (\mathbb{P}_Y) - \mu_{K_d} (\mathbb{P}_X) \otimes \mu_{L_d} (\mathbb{P}_Y)  \right\|_{\mathcal{H}} \\
        & \stackrel{(c)}{=}  \left\| \left[\mu_K (\mathbb{P}_X) - \mu_{K_d} (\mathbb{P}_X) \right] \otimes \mu_L (\mathbb{P}_Y)  \right\|_{\mathcal{H}} + \left\| \mu_{K_d} (\mathbb{P}_X) \left[\mu_L (\mathbb{P}_Y) - \mu_{L_d} (\mathbb{P}_Y) \right]  \right\|_{\mathcal{H}} \\
        & \stackrel{(d)}{\le} \left\|\mu_K (\mathbb{P}_X) - \mu_{K_d} (\mathbb{P}_X) \right\|_{\mathcal{H}_K} \left\|\mu_L (\mathbb{P}_Y)\right\|_{\mathcal{H}_L} + \left\|\mu_{K_d} (\mathbb{P}_X)\right\|_{\mathcal{H}_K} \left\|\mu_L (\mathbb{P}_Y) - \mu_{L_d} (\mathbb{P}_Y) \right\|_{\mathcal{H}_L},
    \end{align*}
    (a) is obtained by adding the term $\pm \mu_{K_d} (\mathbb{P}_X) \otimes \mu_L (\mathbb{P}_Y)$, (b) is followed from triangle inequality, (c) is implied by the bilinear property of tensor product and (d) is because $\left\| S \otimes T \right\| \le \left\| S \right\| \left\| T \right\|$. Once we can deduce the $\left\|\mu_K (\mathbb{P}_X) - \mu_{K_d} (\mathbb{P}_X) \right\|_{\mathcal{H}_K}$ and $\left\|\mu_L (\mathbb{P}_Y) - \mu_{L_d} (\mathbb{P}_Y) \right\|_{\mathcal{H}_L}$ converge to 0 as $n \to \infty$, we can conclude $B_2$ converge to 0.

    Since the deduce process for the convergence of $\left\|\mu_K (\mathbb{P}_X) - \mu_{K_d} (\mathbb{P}_X) \right\|_{\mathcal{H}_K}$ and $\left\|\mu_L (\mathbb{P}_Y) - \mu_{L_d} (\mathbb{P}_Y) \right\|_{\mathcal{H}_L}$, we use $\left\|\mu_K (\mathbb{P}_X) - \mu_{K_d} (\mathbb{P}_X) \right\|_{\mathcal{H}_K}$ as an example. Recalling the definition of kernel mean embedding, we have
    \begin{align*}
        \mu_{K}(\mathbb{P}_{X}) = \mathrm{E}_{\mathbb{P}_{X}} \left[ K(\cdot,x) \right] , \quad \quad \mu_{K_d}(\mathbb{P}_{X}) = \mathrm{E}_{\mathbb{P}_{X}} \left[ K_d(\cdot,x) \right],
    \end{align*}
    then we have
    \begin{align*}
        & \quad \left\|\mu_K (\mathbb{P}_X) - \mu_{K_d} (\mathbb{P}_X) \right\|^2_{\mathcal{H}_K} = \left\| \mathrm{E}_{\mathbb{P}_{X}} \left[ K(\cdot,x) \right] -  \mathrm{E}_{\mathbb{P}_{X}} \left[ K_d(\cdot,x) \right] \right\|^2_{\mathcal{H}_K} \\
        & \stackrel{(a)}{=} \left\| \mathrm{E}_{\mathbb{P}_{X}} \left[ K(\cdot,x) - K_d(\cdot,x) \right] \right\|^2_{\mathcal{H}_K} \stackrel{(b)}{=} \left\| \mathrm{E}_{\mathbb{P}_{X}} \left( I - P_d \right) K(\cdot,x) \right\|^2_{\mathcal{H}_K} \\
        & \stackrel{(c)}{\le} \left( \mathrm{E}_{\mathbb{P}_{X}} \left\| \left( I - P_d \right) K(\cdot,x) \right\|_{\mathcal{H}_K}  \right)^2 \stackrel{(d)}{\le} \mathrm{E}_{\mathbb{P}_{X}} \left\| \left( I - P_d \right) K(\cdot,x) \right\|^2_{\mathcal{H}_K} \\
        & \stackrel{(e)}{=} \left\| (I-P_d)C_K^{1/2}  \right\|^2_{\mathrm{HS}(\mathcal{H}_K,\mathcal{H}_K)} \stackrel{(f)}{\le} 9 \mathcal{N}_{C_K}(t)(\lambda_d + t),
    \end{align*}
    where $C_K := \mathrm{E}_{\mathbb{P}_{X}} \left[ K(\cdot,X) \otimes K(\cdot,X)  \right]$, $\mathrm{HS}(\mathcal{H}_K,\mathcal{H}_K)$ denotes the Hilbert-Schmidt norm and $\lambda_d$ denotes the $d$-th eigenvalue of $C_K$. (a) follows from the property of expectation, (b) is because for kernel $K$ we can find a projection $P_d$ such that $K_d(\cdot,x)=P_d K(\cdot,x)$ as we defined in \textbf{Step 1}, (c) is implied by the homogeneity of norms and the triangle inequality, (d) is obtained by the Jensen’s inequality---$f(\mathrm{E}(Z)) \le \mathrm{E}(f(Z))$ with $f(z)=z^2$, (e) is proved in Equation (13) in \cite{sterge2020gain}. (f) follows from Lemma~\ref{bound_kpca}. Although the original statement of Lemma~\ref{bound_kpca} pertains to the Nystr\"{o}m approximation, the only difference lies in the format of the projection, which does not affect its properties. Therefore, the conclusion in Lemma~\ref{bound_kpca} also holds in our setting.

    Thus we got 
    \begin{align*}
        \left\|\mu_K (\mathbb{P}_X) - \mu_{K_d} (\mathbb{P}_X) \right\|_{\mathcal{H}_K} \le \underbrace{\sqrt{9 \mathcal{N}_{C_K}(t)(\lambda_d + t)}}_{=:r}.
    \end{align*}
    Then we can discuss the convergence condition of $r$ from two decay patterns of eigenvalues of $C_K$:
    \begin{itemize}
        \item When $(\lambda_i)_{i \in I}$ decay polynomially, we have $\lambda_i = \mathcal{O} \left(i^{-\alpha}\right)$, where $\alpha > 1$. Suppose the embedding dimension $d$ is chosen as $d=n^{\theta/ \alpha}$ where $0<\theta \le \alpha$ and we have $\lambda_d = \mathcal{O} \left( n^{-\theta} \right)$. According to (i) of Lemma~\ref{order_Nc}, we have 
        \begin{align*}
            \mathcal{N}_{C_K}(t)=\mathcal{O}\left( t^{-\frac{1}{\alpha}}  \right) \stackrel{(a)}{=} \mathcal{O}\left( \frac{n^{\frac{1}{\alpha}}}{\log^{\frac{1}{\alpha}} (n)}  \right) ,
        \end{align*}
        (a) is because in the statement of Lemma~\ref{bound_kpca} we let $t = \Theta \left(\frac{\log(n)}{n}\right)$. Then we have
        \begin{align*}
            r = \mathcal{O} \left[ \sqrt{\frac{n^{\frac{1}{\alpha}}}{\log^{\frac{1}{\alpha}} (n)} \left( n^{-\theta} + \frac{\log(n)}{n} \right) } \right]  = \left\{ \begin{array}{cl}
         \mathcal{O} \left (\frac{n^{\frac{1}{2\alpha}-1}}{\log^{\frac{1}{2\alpha}-1}(n)} \right),&  \ \theta \geq 1 \\
         \mathcal{O} \left (\frac{n^{\frac{1}{2\alpha}-\frac{\theta}{2}}}{\log ^{\frac{1}{2\alpha}} (n)} \right) ,&  \ \theta < 1
        \end{array} \right.  
        \end{align*}
        Since $\alpha > 1$ and $0<\theta \le \alpha$, thus when $\frac{\theta}{2} > \frac{1}{2\alpha}$, $r$ will converge to 0, which implies that $d = n^{\theta / \alpha} > n^{\frac{1}{\alpha^2}}$.

        \item When $(\lambda_i)_{i \in I}$ decay exponentially, we have $\lambda_i = \mathcal{O} \left(e^{-\tau i}\right)$ for $\tau >0$. Suppose the embedding dimension $d$ is chosen as $d=\frac{1}{\tau} \log \left( n^{\theta} \right)$ for $\theta >0$ and therefore we have $\lambda_d = \mathcal{O}\left( n^{-\theta} \right)$. According to (ii) of Lemma~\ref{order_Nc} we have $\mathcal{N}_{C_K}(t) = \mathcal{O} \left(  \log (1/t)\right) = \mathcal{O} \left(  \log (n/ \log(n))\right) = \mathcal{O} \left(  \log (n) \right)$. Thus we have
        \begin{align*}
            r = \mathcal{O} \left[ \sqrt{\log (n) \left( n^{-\theta} + \frac{\log (n)}{n} \right)}  \right] \to 0 \ \text{as} \ n \to \infty.
        \end{align*}
        
    \end{itemize}

    The above two decay conditions can be seen as sufficient to contain most of the eigen decay conditions of $C$. Therefore, we can conclude that for a specific range of $d$, $\left\|\mu_K (\mathbb{P}_X) - \mu_{K_d} (\mathbb{P}_X) \right\|_{\mathcal{H}_K}$ will converge to 0 as $n \to \infty$.

    For $\left\|\mu_L (\mathbb{P}_Y) - \mu_{L_d} (\mathbb{P}_Y) \right\|_{\mathcal{H}_L}$, we can follow the same discussion to obtain a similar conclusion, thus we solve the convergence of $B_2$.

    \item \textbf{Term $B_1$}: For the $B_1$, since $\mathbb{P}_{XY}$ denote the joint distribution of $X$ and $Y$, by the defnition of kernel mean embedding, we have
    \begin{align*}
       \mu_{G}(\mathbb{P}_{XY}) & = \mathrm{E}_{\mathbb{P}_{XY}} \left[ G(\cdot,x) \right] = \mathrm{E}_{\mathbb{P}_{XY}} \left[ K(\cdot,x) \otimes L(\cdot,y) \right],\\
       \mu_{G_d}(\mathbb{P}_{XY}) & = \mathrm{E}_{\mathbb{P}_{XY}} \left[ G_d(\cdot,x) \right] = \mathrm{E}_{\mathbb{P}_{XY}} \left[ K_d(\cdot,x) \otimes L_d(\cdot,y) \right],
    \end{align*}
    then we have
    \begin{align*}
        B^2_1 & = \left\|  \mathrm{E}_{\mathbb{P}_{XY}} \left[ K(\cdot,x) \otimes L(\cdot,y) \right] - \mathrm{E}_{\mathbb{P}_{XY}} \left[ K_d(\cdot,x) \otimes L_d(\cdot,y) \right]  \right\|^2_{\mathcal{H}} \\
        & = \left\|  \mathrm{E}_{\mathbb{P}_{XY}} \left[ K(\cdot,x) \otimes L(\cdot,y) - K_d(\cdot,x) \otimes L_d(\cdot,y) \right]  \right\|^2_{\mathcal{H}} \\
        & \stackrel{(a)}{\le} \left( \mathrm{E}_{\mathbb{P}_{XY}} \left\|  K(\cdot,x) \otimes L(\cdot,y) - K_d(\cdot,x) \otimes L_d(\cdot,y) \right\|_{\mathcal{H}}  \right)^2 ,
    \end{align*}
    where (a) is implied by the homogeneity of norms and the triangle inequality. Taking the square root of both sides we have
    \begin{align*}
        B_1 & \le \mathrm{E}_{\mathbb{P}_{XY}} \left\|  K(\cdot,x) \otimes L(\cdot,y) - K_d(\cdot,x) \otimes L_d(\cdot,y) \right\|_{\mathcal{H}} \\
        & \stackrel{(a)}{\le} \mathrm{E}_{\mathbb{P}_{XY}} \left\|  K(\cdot,x) \otimes L(\cdot,y) - K_d(\cdot,x) \otimes L(\cdot,y) +  K_d(\cdot,x) \otimes L(\cdot,y) \right. \\
        & \quad \left. - K_d(\cdot,x) \otimes L_d(\cdot,y) \right\|_{\mathcal{H}} \\
        & \stackrel{(b)}{\le} \mathrm{E}_{\mathbb{P}_{XY}} \left\{ \left\|  \left[ K(\cdot,x) - K_d(\cdot,x) \right] \otimes L_d(\cdot,y) \right\|_{\mathcal{H}}  \right\} \\
        & \quad + \mathrm{E}_{\mathbb{P}_{XY}} \left\{ \left\|  \left[ L(\cdot,y) - L_d(\cdot,y) \right] \otimes K_d(\cdot,x) \right\|_{\mathcal{H}} \right\} \\
        & \stackrel{(c)}{\le} \mathrm{E}_{\mathbb{P}_{XY}} \left\{ \left\|  K(\cdot,x) - K_d(\cdot,x) \right\|_{\mathcal{H}_K} \left\| L_d(\cdot,y) \right\|_{\mathcal{H}_L} \right\} \\
        & \quad + \mathrm{E}_{\mathbb{P}_{XY}} \left\{ \left\| K_d(\cdot,x) \right\|_{\mathcal{H}_K} \left\| L(\cdot,y) - L_d(\cdot,y) \right\|_{\mathcal{H}_L} \right\} \\
        & \stackrel{(d)}{\le} \mathrm{E}_{\mathbb{P}_{X}} \left[ \left\|  K(\cdot,x) - K_d(\cdot,x) \right\|_{\mathcal{H}_K} \right] + \mathrm{E}_{\mathbb{P}_{Y}} \left[ \left\|  L(\cdot,y) - L_d(\cdot,y) \right\|_{\mathcal{H}_L} \right],
    \end{align*}
    (a) is obtained by adding the term $K_d(\cdot,x) \otimes L(\cdot,y)$, (b) is by triangle inequality and combining similar terms, (c) is by triangle inequality and (d) is followed from the property that $L(y,{y}')$ and $K(x,{x}')$ are less than 1.

    Therefore we can know the convergence condition for $B_1$ is as same as $B_2$.
    
\end{itemize}

Summarize the conclusions of this step. Let $C_K = \mathrm{E}_{\mathbb{P}_X} \left[ K(\cdot,X) \otimes K(\cdot,X) \right]$ and $C_L = \mathrm{E}_{\mathbb{P}_Y} \left[ L(\cdot,Y) \otimes L(\cdot,Y) \right]$ denote the covariance operator associated with kernel $K$ and $L$, respectively. Let the eigenvalues of $C_K$ and $C_L$ are defined as $(\lambda_i(C_K))_{i \in I}$ and $(\lambda_i(C_L))_{i \in I}$, respectively. For simplicity, we just assume the spectral decay of $C_K$ and $C_L$ are follow the same pattern, and we have the following conclusions:
\begin{itemize}
    \item When the eigenvalues $(\lambda_i(C_K))_{i \in I}$ and $(\lambda_i(C_L))_{i \in I}$ decay polynomially, such that $\lambda_i(C_K) = \mathcal{O}(i^{-\alpha_K})$ and $\lambda_i(C_L) = \mathcal{O}(i^{-\alpha_L})$ for $\alpha_K, \alpha_L >1$, the we should select $d \ge \max \left\{ n^{\frac{1}{\alpha_K^2}}, n^{\frac{1}{\alpha_L^2}}  \right\}$;
    \item When the eigenvalues $(\lambda_i(C_K))_{i \in I}$ and $(\lambda_i(C_L))_{i \in I}$ decay exponentially, such that $\lambda_i(C_K) = \mathcal{O}(e^{-\tau_K i})$ and $\lambda_i(C_L) = \mathcal{O}(e^{-\tau_L i})$ for $\tau_K, \tau_L >0$, by selecting a $\theta > 0$, we can chose $d \ge \max \left\{ \frac{1}{\tau_K} \log(n^{\theta}), \frac{1}{\tau_L} \log(n^{\theta})  \right\}$.
\end{itemize}

\textbf{Step 3. } This step we want to show $\lim_{n \to \infty} \mathrm{gCov}_{n}(G_{1},G_{2})=\mathrm{hCov}^{(d)}(G_{1},G_{2})$. Let $\hat{K}_{ij}=\hat{K}_{d}(x_i,x_j)=\left \langle \hat{\phi}_{d}(x),\hat{\phi}_{d}({x}')  \right \rangle_{L^2(\mathbb{P}_X)}$. Define an inclusion operator $\mathcal{J}:\mathcal{H} \to L^2(\mathbb{P}_X)$, we have:
\begin{align*}
    K_{ij} = K_{d}(x_i,x_j) =\left \langle \mathcal{J} \phi_{d}(x), \mathcal{J} \phi_{d}({x}')  \right \rangle_{L^2(\mathbb{P}_X)}.
\end{align*}
Thus we got
\begin{align}\label{bound_adj}
     & \quad \mathrm{E} \left  | K_{d}(x_{i},x_{j})- \hat{K}_{d}(x_{i},x_{j}) \right | \nonumber\\
     &= \mathrm{E} \left | \left \langle \mathcal{J} \phi_{d}(x_i), \mathcal{J} \phi_{d}(x_j) \right \rangle_{L^2(\mathbb{P}_X)} - \left \langle \hat{\phi}_{d}(x_{i}),\hat{\phi}_{d}(x_{j}) \right \rangle_{L^2(\mathbb{P}_X)} \right | \nonumber \\
     & \stackrel{(a)}{=} \mathrm{E} \left| \left \langle \mathcal{J} \phi_{d}(x_i), \mathcal{J} \phi_{d}(x_j) \right \rangle_{L^2(\mathbb{P}_X)} - \left \langle \mathcal{J} \phi_{d}(x_i), \hat{\phi}_{d}(x_j) \right \rangle_{L^2(\mathbb{P}_X)}  \right. \nonumber\\
     & \quad \left. + \left \langle \mathcal{J} \phi_{d}(x_i), \hat{\phi}_{d}(x_j) \right \rangle_{L^2(\mathbb{P}_X)} -\left \langle \hat{\phi}_{d}(x_{i}),\hat{\phi}_{d}(x_{j}) \right \rangle_{L^2(\mathbb{P}_X)} \right|  \nonumber \\
     & \stackrel{(b)}{\le} \mathrm{E} \left| \left \langle \mathcal{J} \phi_{d}(x_i), \mathcal{J} \phi_{d}(x_j)- \hat{\phi}_{d}(x_{j}) \right \rangle_{L^2(\mathbb{P}_X)}  \right| \nonumber\\
     & \quad + \mathrm{E} \left| \left \langle \mathcal{J} \phi_{d}(x_i)- \hat{\phi}_{d}(x_i) , \hat{\phi}_d (x_j) \right \rangle_{L^2(\mathbb{P}_X)}  \right| \nonumber \\
     &\le \mathrm{E}  \left \{  \left \| \mathcal{J} \phi_{d}(x_i) \right \|_2 \left \| \mathcal{J} \phi_{d}(x_j)- \hat{\phi}_{d}(x_{j}) \right \|_2 \right \} + \mathrm{E}  \left \{ \left \| \mathcal{J} \phi_{d}(x_i)- \hat{\phi}_{d}(x_i)  \right \|_2 \left \|  \hat{\phi}_d (x_j) \right \|_2  \right \} \nonumber \\
     & \stackrel{(c)}{\le} C(n) 54 \delta _{d}^{-2}\sqrt{\frac{6d \log n}{n}}, 
\end{align}
(a) is obtained by adding $\left \langle \mathcal{J} \phi_{d}(x_i), \hat{\phi}_{d}(x_j) \right \rangle_{L^2(\mathbb{P}_X)}$, (b) is implied by triangle inequality, (c) is obtained by the \eqref{bound_adj_appro} in Lemma~\ref{concentration_adj} and $\left \| \mathcal{J} \phi_{d}(x_i) \right \|_2, \left \|  \hat{\phi}_d (x_j) \right \|_2 \le 1$. It is worth noting that our $\hat{\Phi}_d$ is calculated by the spectral decomposition of $(\A^\top \A)^{1/2}$, while the Lemma~\ref{bound_psd_trans} shows that the spectral norm of $(\A^\top \A)^{1/2} - \mathbf{K}$ is almost Lipschitz-continuous, and the order of $C(n)$ is of $\log(n)$, therefore the convergence performance of Lemma~\ref{concentration_adj} remains unchanged.

After we obtained the difference between the single kernel, we use the decomposition in \cite{szekely2014partial} such that
\begin{equation*}
n(n-3)\mathrm{gCov}_n(G_1,G_2)= T_1 - \frac{T_2}{(n-1)(n-2)^2}-\frac{2T_3}{n-2},
\end{equation*}
where $T_1 =  {\textstyle \sum_{i\ne j}}\hat{K}_{ij}\hat{L}_{ij} $, $T_2=\hat{K}_{\cdot \cdot}\hat{L}_{\cdot \cdot} $ and $T_3 =  {\textstyle \sum_{i=1}^n}\hat{K}_{i \cdot}\hat{L}_{i \cdot} $. For the first item $T_1$ we have:
\begin{equation*}
\begin{aligned}
& \quad \frac{1}{n(n-3)}\left \{ E(T_1) - E({\textstyle \sum_{i\ne j}}K_{ij} L_{ij} ) \right \} =  \frac{1}{n(n-3)} \sum_{i \ne j} E(\hat{K}_{ij}\hat{L}_{ij}  - K_{ij} L_{ij}) \\
& = \frac{1}{n(n-3)} \sum_{i \ne j} E(\hat{K}_{ij}\hat{L}_{ij}  - \hat{K}_{ij} L_{ij} + \hat{K}_{ij} L_{ij} - K_{ij} L_{ij}) \\ 
& = \frac{1}{n(n-3)} \sum_{i \ne j} E \left \{ \hat{K}_{ij}(\hat{L}_{ij}- L_{ij}) + L_{ij}(\hat{K}_{ij}-K_{ij})\right \} \\
& \le \frac{1}{n(n-3)} \sum_{i \ne j} \left\{ E(\hat{L}_{ij}-L_{ij}) + E(\hat{K}_{ij}-K_{ij})   \right \} \to 0 \quad \text{as} \quad n \to \infty.
\end{aligned}
\end{equation*}
The last inequality is because $K_{ij}$ is a probability under our assumption so we have $K_{ij} \le 1$. The bound for $T_2$ and $T_3$ are similar. So that we have 
\begin{equation*}
E\left \{ \mathrm{gCov}_{n}(G_{1},G_{2}) \right \}-\mathrm{hCov}^{(d)}(G_{1},G_{2}) \to 0 \quad \text{as} \quad n \to \infty.
\end{equation*}
which also means the proposed $\mathrm{gCov}_{n}(G_{1},G_{2})$ can approximate the d-th truncation of $\mathrm{hCov}(G_{1},G_{2})$, where random graphs are generated by $\X$ and $\Y$.

Besides, from the error bound in \eqref{bound_adj}, we can know $d = \mathcal{O}\left( \frac{n}{\log n} \right)$ ensures that 
$$\lim_{n \to \infty} \mathrm{gCov}_{n}(G_{1},G_{2}) \to \mathrm{hCov}(G_{1},G_{2}).$$


\end{proof}

\begin{proof}[Proof of Theorem 2]

When $G_{1}$ and $G_{2}$ are independent, the random variables $\left \{ X_{i} \right \}_{i=1}^n$ and $\left \{ Y_{i} \right \}_{i=1}^n$ that generate $G_{1}$ and $G_{2}$ are independent. Considering the Hilbert-Schmidt Independent Criterion (HISC) of random variables X and Y. The Hilbert-Schmidt Independence Criterion is given by $\mathrm{HSIC}(\mathbb{P}_{XY},\mathcal{H}_K,\mathcal{H}_L):= \left \| C_{XY} \right \|_{\mathcal{H}}^2$,
where $\mathbb{P}_{XY}$ denotes the joint measure, $\mathcal{H}_K$ and $\mathcal{H}_L$ are RKHSs associated with $X$ and $Y$. $C_{XY}$ denotes the cross-covariance operator. 
Theorem 4 in \citep{gretton2005measuring} shows $\left \| C_{XY} \right \|_{\mathcal{H}}=0$ if and only if $X$ and $Y$ are independent. 
Then the independent criterion is equivalent to $\mathrm{hCov}(G_{1},G_{2})=0$ if and only if $G_{1}$ and $G_{2}$ are independent. In the Theorem 1 we proved that $\mathrm{gCov}_{n}(G_{1},G_{2})$ converge to $\mathrm{gCov}(G_{1},G_{2})$ almost surely. 

Besides, when $G_{1}$ and $G_{2}$ are independent we have $\mathrm{hCov}(G_{1},G_{1})=0$ and $\mathrm{hCor}(G_{1},G_{1})=0$. On the other hand, $\mathrm{hCor}(G_{1},G_{1})>0$ if dependent. Then follows the convergence property of $\mathrm{gCor}_{n}(G_{1},G_{1})$, the proof was completed.

\end{proof}

Before the proof of Theorem 3, we elaborate some details about the proposed permutation test. Let $\varphi:[n] \to [n]$ denote a permutation of $[n]$. Recall that in our estimator we have $\hat{\K}=\hat{\bm{\Phi}}_{1,d} \hat{\bm{\Phi}}_{1,d}^\top$ where $\hat{\bm{\Phi}}_{1,d} = \left( \hat{\phi}_{1,d}(\x_1),\dots,\hat{\phi}_{1,d}(\x_n)  \right)^\top \in \R^{n \times d}$. Then we can shuffle the rows of $\hat{\bm{\Phi}}_{1,d}$ according to $\varphi$ such that
\begin{align*}
     \left( \hat{\phi}_{1,d}(\x_1),\dots,\hat{\phi}_{1,d}(\x_n)  \right)^\top \overset{\varphi}{\rightarrow}  \left( \hat{\phi}_{1,d}(\x_{\varphi(1)}),\dots,\hat{\phi}_{1,d}(\x_{\varphi(n)}) \right)^\top.
\end{align*}
The shuffled $\hat{\bm{\Phi}}_{1,d}$ is denoted as $\hat{\bm{\Phi}}^*_{1,d}$. Repeating the process $B$ times and by using the $\hat{\bm{\Phi}}^*_{1,d}$ obtained by each time we can recalculate the $\mathrm{gCor}_n(G_1,G_2)$ and denote it as $\mathrm{gCor}_n^{\varphi_b}(G_1,G_2)$ for $b=1,\dots,B$. The original calculated $\mathrm{gCor}_n(G_1,G_2)$ is denoted as $\mathrm{gCor}^0_n(G_1,G_2)$ and we can define our decision rule as follows:
\begin{align*}
    f_n \defn \mathbbm{1}\left \{ \mathrm{gCor}^0_n(G_1,G_2) \ge c_{n,1-\alpha} \right \},
\end{align*}
where the permutation critical value $c_{n,1-\alpha}$ denotes the $(1-\alpha)$-quantile of the permutation distribution.

The following proof shows that our proposed test is 
\begin{itemize}
    \item[(i)] under the alternative hypothesis, $\lim_{n\to\infty}\mathbb{P}(f_n=1)=1$;
    \item[(ii)] under $H_0$, the test is asymptotically valid with level at most $\alpha$, in other words, $\limsup_{n\to\infty}\sup \mathbb{P}(f_n=1)\le\alpha$.
\end{itemize}
Therefore the sample method is universally consistent. Besides, we also proved for the finte-sample, the permutation test is asymptotic valid.

\begin{proof}[Proof of Theorem 3]

We define $\hat{R}(t)$ as:
\begin{equation*}
\hat{R}(t) \defn \frac{1}{B}\sum_{b=1}^{B}\mathbbm{1}\left \{ \mathrm{gCor}_n^{\varphi_b}(G_1,G_2) \le  t \right \} \in [0,1] , \quad  t\in \mathbb{R}.
\end{equation*}
Then for a level $\alpha \in (0,1)$, let $\hat{R}^{-1}(1-\alpha)$ be the $(1-\alpha)$ quantile of $\hat{R}$. Then we reject the null hypothesis iff $\mathrm{gCor}^0_n(G_1,G_2) \ge \hat{R}^{-1}(1-\alpha)$.
From the definition of $f_n$, $f_n=1$ means that $\hat{R}^{-1}(1-\alpha) \le \mathrm{gCor}^0_n(G_1,G_2)$, which holds if and only if 
    \begin{align*}
        1-\alpha \le \frac{1}{B} \sum_{b=1}^B \mathbbm{1}_{ \left \{ \mathrm{gCor}_n^{\varphi_b}(G_1,G_2) \le \mathrm{gCor}_n^0(G_1,G_2) \right \}  } \Leftrightarrow \alpha \ge \frac{1}{B} \sum_{b=1}^B \mathbbm{1}_{ \left \{ \mathrm{gCor}_n^{\varphi_b}(G_1,G_2) \ge \mathrm{gCor}_n^0(G_1,G_2) \right \}  }.
    \end{align*}
Given the LPRGs $G_1$ and $G_2$, we define the $p_G$ as follows:
    \begin{align*}
    \pr(f_n=1) =: p_G = \pr \left\{ \mathrm{gCor}_n^{\varphi}(G_1,G_2) \ge \mathrm{gCor}_n^0(G_1,G_2) \mid G_1,G_2 \right\} ,
    \end{align*}
and we assume $p_G$ contains all permutations of $\hat{\bm{\Phi}}_{1,d}$. Then the alternative hypothesis holds when 
\begin{align*}
    p_G = \frac{1}{n!} \sum_{b=1}^{n!} \mathbbm{1}_{ \left \{ \mathrm{gCor}_n^{\varphi_b}(G_1,G_2) \ge \mathrm{gCor}_n^0(G_1,G_2) \right \}  }  \le \alpha
\end{align*}
Let $r = \left\lfloor \alpha n! \right\rfloor$, where $\left\lfloor a \right\rfloor$ denote the largest integer less than or equal to $a$ and we have $r/n! \to \alpha$. Let $t^r_n(G_1,G_2)$ be the value of $\mathrm{gCor}_n^{\varphi}(G_1,G_2)$ computed by a kind of permutation and $t_n(G_1,G_2)$ denote the $\mathrm{gCor}_n^0(G_1,G_2)$. In particular, we reject $H_0$ when $t_n(G_1,G_2) \ge t^r_n(G_1,G_2)$, therefore we have
\begin{align*}
    \pr (p_G \le \alpha) \ge \pr (t_n(G_1,G_2) \ge t^r_n(G_1,G_2)).
\end{align*}
Notice for any $\epsilon > 0$, $t^r_n(G_1,G_2) \ge \epsilon$ implies that
\begin{align*}
    \pr \left\{ \mathrm{gCor}_n^{\varphi}(G_1,G_2) \ge \epsilon \mid G_1,G_2  \right\} \ge r/n!.
\end{align*}
Therefore we have
\begin{align*}
    \lim_{n \to \infty} \pr (t^r_n(G_1,G_2) \ge \epsilon) & \le \lim_{n \to \infty} \pr \left\{ \pr \left( \mathrm{gCor}_n^{\varphi}(G_1,G_2) \ge \epsilon \mid G_1,G_2  \right) \ge r/n! \right\} \\
    & \overset{(a)}{\le} \lim_{n \to \infty} \frac{E \left\{ \pr \left( \mathrm{gCor}_n^{\varphi}(G_1,G_2) \ge \epsilon \mid G_1,G_2  \right) \right\}}{r/n!} \\
    & = \lim_{n \to \infty} \frac{ \pr \left( \mathrm{gCor}_n^{\varphi}(G_1,G_2) \ge \epsilon \right)}{r/n!}=0,
\end{align*}
where $(a)$ is obtained by the Markov's inequality. The last equality is because we have $\mathrm{gCor}_n^{\varphi}(G_1,G_2)$ converge to 0 by Theorem 2 and $r/n! \to \alpha$. Then we have
\begin{align*}
    \lim_{n \to \infty} \pr(f_n=1) = \lim_{n \to \infty}\pr(p_G \le \alpha) & \ge \lim_{n \to \infty} \pr (t_n(G) \ge t^r_n(G)) \\
    &= \lim_{n \to \infty} \left( 1-\pr (t_n(G) < t^r_n(G)) \right) =1
\end{align*}
under alternative hypothesis.

Under the null hypothesis, we have $\mathrm{gCor}_n^0(G_1,G_2)$ converge to 0 in probability, therefore $\mathrm{gCor}_n^0(G_1,G_2)$ has same asymptotic distribution as $\mathrm{gCor}_n^{\varphi}(G_1,G_2)$. Then we have
\begin{align*}
    \lim_{n \to \infty} \hat{R}(t) & = \lim_{n \to \infty} \frac{1}{n!} \sum_{b=1}^{n!} \mathbbm{1}_{ \left \{\mathrm{gCor}_n^{\varphi}(G_1,G_2) \le t \right \}  }  = \lim_{n \to \infty} \pr \left( \mathrm{gCor}_n^{\varphi}(G_1,G_2) \le t \right) =: G(t).
\end{align*}
Since $G(t)$ is continuous in $(0,1)$ then we have $\lim_{n \to \infty} \hat{R}^{-1}(t) = G^{-1}(t)$. Therefore
\begin{align*}
    & \quad \lim_{n \to \infty} \sup \pr (f_n=1) = \lim_{n \to \infty} \sup \pr \left( \mathrm{gCor}_n^0(G_1,G_2) \ge \hat{R}^{-1}(1-\alpha) \right) \\
    & = 1- \lim_{n \to \infty}\inf \pr \left( \mathrm{gCor}_n^0(G_1,G_2) \le \hat{R}^{-1}(1-\alpha) \right) = 1- G(G^{-1}(1-\alpha) = \alpha,
\end{align*}
thus we have $\pr(f_n=1) \le \alpha$ under $H_0$. Combine the above results we can know the permutation test is consistent under population level. 

Under the finite sample we have
\begin{align*}
    \hat{p} =  \frac{1+ \sum_{b=1}^B \mathbbm{1} \left\{ \mathrm{gCor}_n^{\varphi}(G_1,G_2) \ge \mathrm{gCor}_n^0(G_1,G_2)  \right\}}{B+1} ,
\end{align*}
where $B$ ia a sufficient large number. Let $R \defn \mathbbm{1} \left\{ \mathrm{gCor}_n^{\varphi}(G_1,G_2) \ge \mathrm{gCor}_n^0(G_1,G_2)  \right\}$ and $R \mid D \sim \text{Binom}(B,p_G)$. 
Considering for any probability such that null hypothesis holds, the finite sample permutation test is consistent when $\lim_{n \to \infty} \pr (\hat{p} \le \alpha)=1$.

For the finite-sample permutation test, since $R \mid D \sim \text{Binom}(B,p_G)$ and $p_G \to 0$ according to the above discussion, then we chose $\epsilon_1,\epsilon_2 \ge 0$ and we can have $\pr(p_G \le \epsilon_1) \ge 1-\epsilon_2$, therefore
\begin{align*}
    \pr (\hat{p} \le \alpha) \ge \pr(\hat{p}=\frac{1}{B+1} \mid p_G \le \epsilon_1) \pr(p_G \le \epsilon_1 ) \ge (1-\epsilon_1)^B (1-\epsilon_2) \to 1.
\end{align*}
Then the finite-sample permutation test also consistent.

\end{proof}

\begin{proof}[Proof of Theorem 4]

The unbiased estimator of sample distance covariance can be calculated by the following formula:
\begin{equation*}
\mathrm{dCov}_{n}(X,Y)=\left ( \tilde{\mathbf{B}}\cdot \tilde{\mathbf{C}}  \right ) =\frac{1}{n(n-3)} \sum_{i\ne j}\tilde{B}_{ij} \tilde{C}_{ij},
\end{equation*}
where $\tilde{\mathbf{B}}$ and $\tilde{\mathbf{C}}$ are $\mathcal{U}$-centered matrices of the distance matrices $\mathbf{B}$ and $\mathbf{C}$. Considering the notation $B_{k\cdot}=\sum_{l=1}^n B_{kl}$, $B_{\cdot l}=\sum_{k=1}^n B_{kl}$ and $B_{\cdot \cdot}=\sum_{k,l=1}^n B_{kl}$. The Proposition 1 in \citep{szekely2014partial} shows that the $\left ( \tilde{\mathbf{B}}\cdot \tilde{\mathbf{C}}  \right )$ can be rewritten as:
\begin{equation*}
n(n-3)\left ( \tilde{\mathbf{B}}\cdot \tilde{\mathbf{C}}  \right ) = T_{1}-\frac{T_{2}}{(n-1)(n-2)^2}-\frac{2T_{3}}{n-2},
\end{equation*} 
where 
\begin{equation*}
T_{1}=\sum_{k \ne l}B_{kl}C_{kl}, \quad T_{2}=B_{\cdot \cdot} C_{\cdot \cdot}, \quad T_{3}=\sum_{k}B_{k \cdot}C_{k \cdot}.
\end{equation*}
Let 
\begin{equation*}
\begin{aligned}
n(n-3)\mathrm{dCov}_{n}(\tilde{\X},\tilde{\Y})=\tilde{T}_{1}-\frac{\tilde{T}_{2}}{(n-1)(n-2)^2}-\frac{2\tilde{T}_{3}}{n-2},  \\
n(n-3)\mathrm{dCov}_{n}(\breve{\X},\breve{\Y})=\breve{T}_{1}-\frac{\breve{T}_{2}}{(n-1)(n-2)^2}-\frac{2\breve{T}_{3}}{n-2}.
\end{aligned}
\end{equation*}
The sepctral decomposition of $\L(\A_{1})$ is $\breve{\X}$ and $\L(\A_{2})$ is $\breve{\Y}$, and the spectral decomposition of $\L(\K_{1})$ is $\tilde{\X}$ and $\L(\K_{2})$ is $\tilde{\Y}$. let $\breve{X}_{i}$ the i-th row of $\breve{\X}$, $\breve{Y}_{i}$ the i-th row of $\breve{\Y}$, $\tilde{X}_{i}$ the i-th row of $\tilde{\X}$ and $\tilde{Y}_{i}$ the i-th row of $\tilde{\Y}$. Then we have
\begin{equation*}
\begin{aligned}
\tilde{T}_{1} - \breve{T}_{1} & = \sum_{k \ne l}\left ( \left \| \tilde{X}_{k}-\tilde{X}_{l} \right \|_2 \left \| \tilde{Y}_{k}-\tilde{Y}_{l} \right \|_2 - \left \| \breve{X}_{k}-\breve{X}_{l} \right \|_2 \left \| \breve{Y}_{k}-\breve{Y}_{l} \right \|_2  \right )  \\
& = \sum_{k \ne l} \left( \left \| \tilde{X}_{k}-\tilde{X}_{l} \right \|_2 \left \| \tilde{Y}_{k}-\tilde{Y}_{l} \right \|_2 -\left \| \tilde{X}_{k}-\tilde{X}_{l} \right \|_2 \left \| \breve{Y}_{k}-\breve{Y}_{l} \right \|_2 \right.\\
& \quad \left. + \left \| \tilde{X}_{k}-\tilde{X}_{l} \right \|_2 \left \| \breve{Y}_{k}-\breve{Y}_{l} \right \|_2 - \left \| \breve{X}_{k}-\breve{X}_{l} \right \|_2 \left \| \breve{Y}_{k}-\breve{Y}_{l} \right \|_2 \right) \\ 
& = l_{1}+l_{2}
\end{aligned}
\end{equation*}
where 
\begin{equation*}
\begin{aligned}
l_{1} &= \sum_{k \ne l}\left ( \left \| \tilde{X}_{k}-\tilde{X}_{l} \right \|_2 \left \| \tilde{Y}_{k}-\tilde{Y}_{l} \right \|_2 -\left \| \tilde{X}_{k}-\tilde{X}_{l} \right \|_2 \left \| \breve{Y}_{k}-\breve{Y}_{l} \right \|_2  \right )  \\
l_{2} &=\sum_{k \ne l}\left ( \left \| \tilde{X}_{k}-\tilde{X}_{l} \right \|_2 \left \| \breve{Y}_{k}-\breve{Y}_{l} \right \|_2 - \left \| \breve{X}_{k}-\breve{X}_{l} \right \|_2 \left \| \breve{Y}_{k}-\breve{Y}_{l} \right \|_2  \right ). 
\end{aligned}
\end{equation*}
Then we have
\begin{align*}
    \left| l_1 \right| & = \left| \sum_{k \ne l} \left \| \tilde{X}_{k}-\tilde{X}_{l} \right \|_2 \left( \left \| \tilde{Y}_{k}-\tilde{Y}_{l} \right \|_2 - \left \| \breve{Y}_{k}-\breve{Y}_{l} \right \|_2  \right )  \right|\\
    & \stackrel{(a)}{\le} \sum_{k \ne l} \left \| \tilde{X}_{k}-\tilde{X}_{l} \right \|_2 \left|  \left \| \tilde{Y}_{k}-\tilde{Y}_{l} \right \|_2 - \left \| \breve{Y}_{k}-\breve{Y}_{l} \right \|_2  \right| \\
    & \stackrel{(b)}{\le} \sum_{k \ne l} \left \| \tilde{X}_{k}-\tilde{X}_{l} \right \|_2 \left \| \tilde{Y}_{k}-\tilde{Y}_{l} -\breve{Y}_{k} + \breve{Y}_{l} \right \|_2 \\
    & \stackrel{(c)}{\le} \sum_{k \ne l} \left \| \tilde{X}_{k}-\tilde{X}_{l} \right \|_2 \left( \left \| \tilde{Y}_{k} -\breve{Y}_{k}\right \|_2 + \left \| \breve{Y}_{l} - \tilde{Y}_{l} \right \|_2  \right) ,
\end{align*}
(a) is implied by $\left | a+b   \right| \le \left | a \right| + \left | b   \right|$, (b) is followed from reverse triangle inequality and (c) is followed from triangle inequality. Therefore by using the results in Lemma~\ref{concentration_laplace} we can have $l_1 \to 0$ as $n \to \infty$, and similarly for $l_2$. The $T_{2}$ and $T_{3}$ can also be proved similarly.

\end{proof}

\begin{proof}[Proof of Theorem 5]

For the semi-sparse random graph, we denote the concentration property as 
\begin{align*}
  \left \|  \A - E\A \right \| \le C
\end{align*}
holds with probability at least $1-\eta$ for $\eta \in (0,1)$. Since $\mathrm{E} \A = \rho_{n} \bP$, then we have
\begin{align}\label{sparse_concentrate}
  \left \|  \tilde{\A} - \bP \right \| \le \rho_{n}^{-1} C
\end{align}
where $\tilde{\A} = \rho_{n}^{-1} \A$. Let the eigendecomposition of $\bP$ is $\U_{P,d}\bS_{P,d}\U_{P,d}^{\top}$, where $\U_{P.d}$ composed by the eigenvectors corresponding to the $d$ largest eigenvalues. Let $\mathcal{P}_{\tilde{\A}}=\U_{\tilde{A},d}\U_{\tilde{A},d}^{\top}$ and $\mathcal{P}_{\bP}=\U_{P,d}\U_{P,d}^{\top}$ and $\delta_{d}=\lambda_{d}-\lambda_{d+1}$, where $(\lambda_i)_{i \in I}$ are non-increasing eigenvalues of integral operator $T_K$. The Theorem B.2 in \citep{tang2013universally} shows that with probability at least $1-\eta$, we have
\begin{equation*}
    \frac{\lambda_{d}(\bP)}{n} - \frac{\lambda_{d+1}(\bP)}{n} \ge \delta_{d}-4\sqrt{2}\sqrt{\frac{\log{(2/\eta)} }{n} }.
\end{equation*}
Meanwhile, define the sets $S_{1}$ and $S_{2}$ as follows:
\begin{align*}
    S_{1}=\left \{ \lambda:\lambda \ge n \lambda_{d}(\bP)-\rho_{n}^{-1}C\right \}, \quad
    S_{2}=\left \{ \lambda:\lambda <  n \lambda_{d+1}(\bP)+\rho_{n}^{-1}C\right \}.
\end{align*}
Then
\begin{equation*}
    \mathrm{dist}(S_{1},S_{2}) \ge n\delta_{d}-4\sqrt{2}\sqrt{n\log{(2/\eta)} } -2\rho_{n}^{-1}C.
\end{equation*}
Assume $S_{1}$ and $S_{2}$ are disjoint, then we have $\mathrm{dist}(S_{1},S_{2})>0$. According to the $\sin \Theta $ theorem in \citep{davis1970rotation} we have:
\begin{equation*}
   \begin{aligned}
   \left \| \mathcal{P}_{\tilde{\A}}(S_{1})-\mathcal{P}_{\bP}(S_{1}) \right \| & \le \frac{\left \| \tilde{\A}-\bP \right \|}{\mathrm{dist}(S_{1},S_{2})} \le \frac{\rho_{n}^{-1}C}{n\delta_{d}-4\sqrt{2}\sqrt{n\log{(2/\eta)} } -2\rho_{n}^{-1}C} \le \frac{2\rho_{n}^{-1}C}{n\delta_{d}}
   \end{aligned}
\end{equation*}
Since $4\sqrt{2} \sqrt{n\log{(2/\eta)}} + 2\rho_{n}^{-1}C \le n \delta_{d}/2$. Then we have
\begin{equation*}
    \begin{aligned}
    \left \| \mathcal{P}_{\tilde{\A}}(\tilde{\A})-\mathcal{P}_{\bP}(\bP) \right \| & \le \left \| \mathcal{P}_{\tilde{\A}}(\tilde{\A}-\bP)  \right \| +\left \| (\mathcal{P}_{\tilde{\A}}-\mathcal{P}_{\bP})\bP \right \|  \\
    & \le \rho_{n}^{-1} C + \frac{2\rho_{n}^{-1}C}{\delta_{d}} \le \delta_{d}^{-1} \rho_{n}^{-1} C + 2\delta_{d}^{-1} \rho_{n}^{-1}C = 3 \delta_{d}^{-1}\rho_{n}^{-1} C.
    \end{aligned}
\end{equation*}
Having $\mathcal{P}_{\tilde{\A}}=\U_{\tilde{A},d}\U_{\tilde{A},d}^{\top}=\rho_{n}^{-1}\U_{A,d}\U_{A,d}^{\top}=\rho_{n}^{-1}\mathcal{P}_{\A}$, for some orthogonal matrix $\mathbf{W} \in \mathbb{R}^{d \times d}  $, we can obtain the following inequalities:
\begin{equation*}
\begin{aligned}
    & \quad \left \| \rho_{n}^{-1/2}\U_{A,d}\bS_{A,d}^{1/2} \W -\U_{P,d}\bS_{P,d}^{1/2} \right \|_F \le \left \| \mathcal{P}_{\tilde{\A}}(\tilde{\A})-\mathcal{P}_{\bP}(\bP) \right \|\frac{\sqrt{d\mathcal{P}_{\tilde{\A}}(\tilde{\A})}+\sqrt{d \mathcal{P}_{\bP}(\bP)}}{\lambda_{d}(\bP)} \\
    & \le 3 \delta_{d}^{-1}\rho_{n}^{-1} C \frac{2\sqrt{dn} }{\lambda_{d}(\bP)} \le 3 \delta_{d}^{-1}\rho_{n}^{-1} C \frac{4\sqrt{dn} }{\lambda_{d}} \le 12 \delta_{d}^{-2}\rho_{n}^{-1} C \sqrt{\frac{d}{n}}.
\end{aligned}
\end{equation*}
Let $\hat{\phi}_{d}(x_{i})$ denotes the i-th column of $\rho_{n}^{-1/2}\U_{A,d}\bS_{A,d}^{1/2}\W$ and $\phi_{d}(x_{i})$ denotes the i-th column of $\U_{P,d}\bS_{P,d}^{1/2}$. Then let $\eta=n^{-2}$ we have
\begin{equation*}
    \begin{aligned}
    \mathrm{E} \left [ \left \| \hat{\phi}_{d}(x_{i}) -\phi_{d}(x_{i}) \right \|  \right ] & \le \sqrt{ \mathrm{E} \left [ \left \| \hat{\phi}_{d}(x_{i}) -\phi_{d}(x_{i}) \right \| ^2 \right ] }  \\
    & \le \sqrt{\frac{1}{n} \mathrm{E} \left [ \left \| \rho_{n}^{-1/2}\U_{A,d}\bS_{A,d}^{1/2}\W -\U_{P,d}\bS_{P,d}^{1/2} \right \|_{F} ^2 \right ] } \\
    & \le \frac{1}{\sqrt{n} }\sqrt{(1-\frac{2}{n^2} )\left ( 12 \delta_{d}^{-2}\rho_{n}^{-1} C \sqrt{\frac{d}{n}} \right )^2 +\frac{2}{n^2}2n }  \\ 
    & \le 12 \delta_{d}^{-2}\rho_{n}^{-1} C\frac{\sqrt{2d}}{n} .
    \end{aligned}
\end{equation*}
The condition (C1) shows that $n \rho_n = \omega \left\{ (\log n)^{\frac{1}{2}+\epsilon}   \right\}$ for $\epsilon>0$ and condition (C2) implies that $C=\sqrt{\log n}$. Thus we have $\rho^{-1} = \mathcal{O}\left( \frac{n}{\left(  \log n\right)^{\frac{1}{2}+\epsilon}} \right)$, by $\epsilon>0$ we can have the $12 \delta_{d}^{-2}\rho_{n}^{-1} C\frac{\sqrt{2d}}{n} = \mathcal{O}\left\{1/ (\log n)^{\epsilon} \right\} \to 0$ as $n \to \infty$ when $d \le c_1 \left( \log n  \right)^{2 \epsilon}$ for some constant $c_1 >0$.

Then, by combining the proof process of Theorem 1, we can derive the following:
\begin{gather*}
   \mathrm{gCov}_{n}(G_{1},G_{2}) \overset{n\rightarrow \infty }{\longrightarrow}  \mathrm{gCov}(G_{1},G_{2}).
\end{gather*}

\end{proof}

\section{Numerical Simulation}\label{sec:additional_sim}

\begin{figure}
        		\scriptsize
        		\begin{center}
        			\begin{tabular}{c}
        				\includegraphics[width=1.0\textwidth]{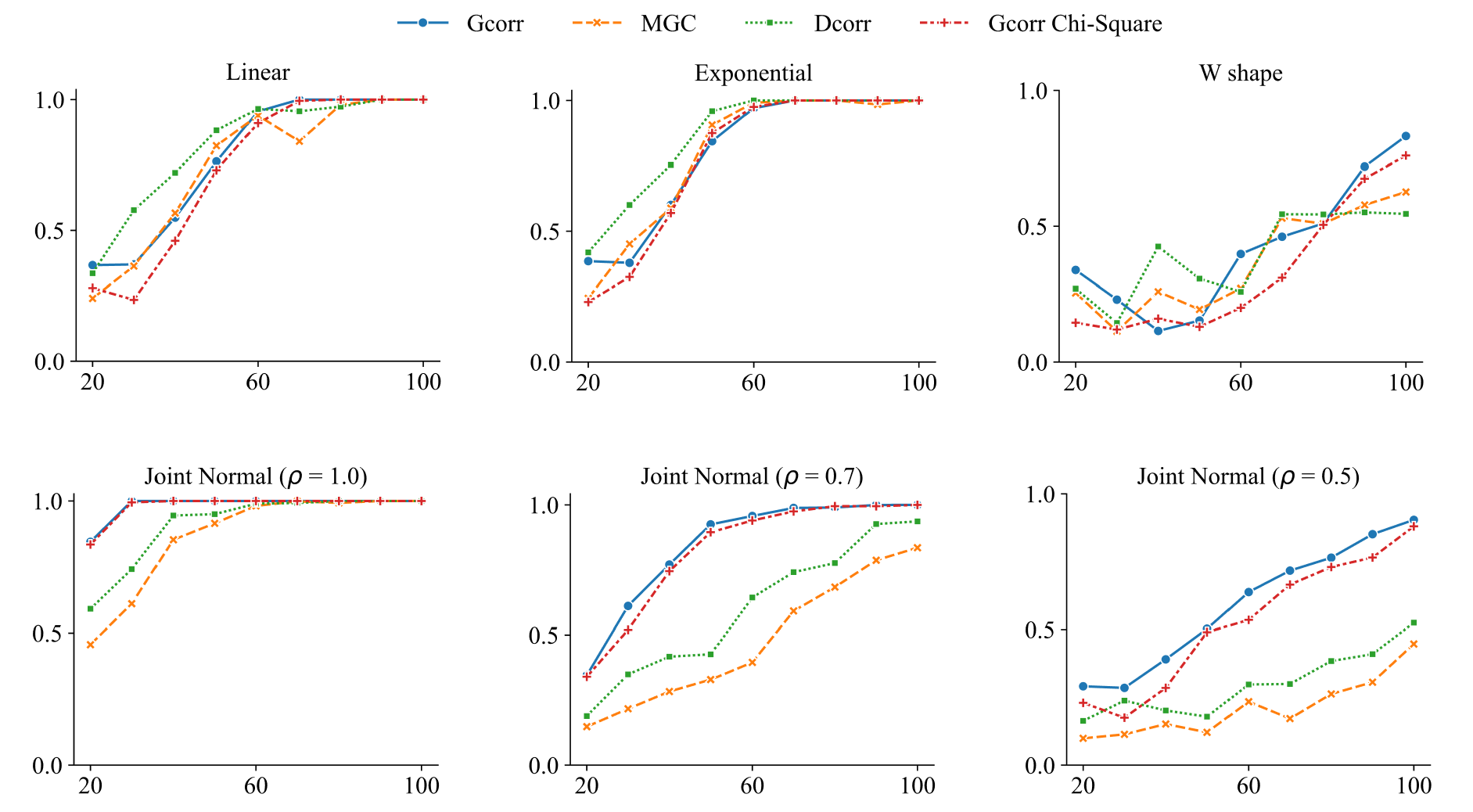}
        				\\
        			\end{tabular}
        		\end{center}
        		\caption{Comparing the testing power of Gcorr (permutation), MGC, Dcorr and Gcorr (Chi-Square) for univariate simulations.}
        		\label{fig:compare_chi_uni_pow}
        \end{figure}

\begin{figure}
        		\scriptsize
        		\begin{center}
        			\begin{tabular}{c}
        				\includegraphics[width=1.0\textwidth]{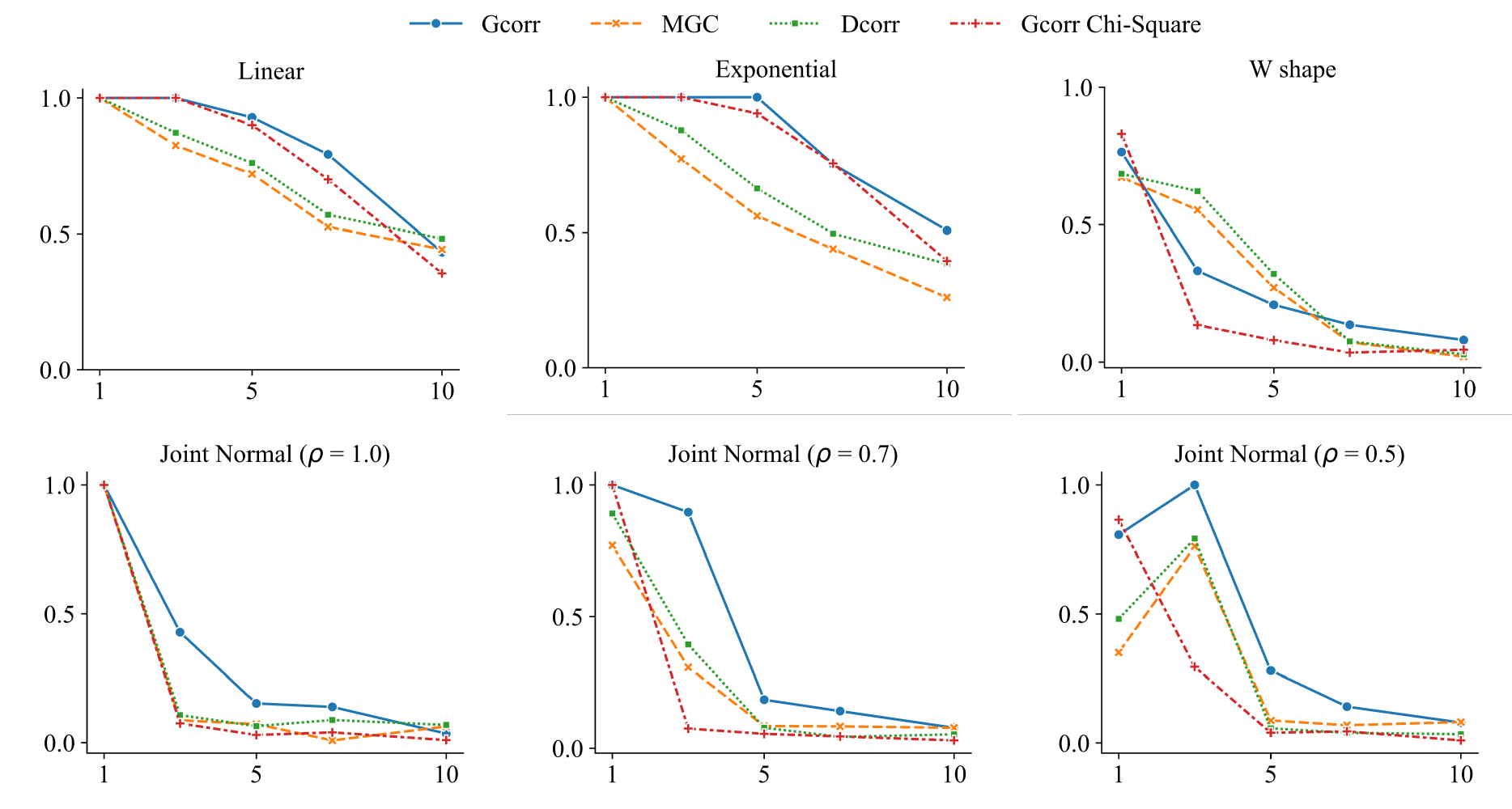}
        				\\
        			\end{tabular}
        		\end{center}
        		\caption{Comparing the testing power of Gcorr (permutation), MGC, Dcorr and Gcorr (Chi-Square) for multivariate simulations.}
        		\label{fig:compare_chi_multi_pow}
        \end{figure}

This section provides additional results from the numerical simulations. We first investigated the chi–square test proposed by \citet{shen2022chi} as a fast alternative to permutation testing. Following their results, for the sample correlation statistic $C = \mathrm{gCor}_n(G_1, G_2)$, the $p$-value can be calculate as $p=\mathbb{P} (nC < \X_1^2 -1)$. We compared the resulting testing power with that obtained via permutation in both univariate and multivariate latent position settings (Figures~\ref{fig:compare_chi_uni_pow} and \ref{fig:compare_chi_multi_pow}). In the univariate case, the chi–square method yields power virtually identical to the permutation test, confirming the validity in low-dimensional settings. In the univariate case, the chi–square method yields power virtually identical to the permutation test, confirming the validity of the approximation in low dimensional settings. However, in the multivariate setting, particularly under joint normal latent positions, we observe a noticeable loss of power as the feature dimension increases. Consequently, while the chi–square test can serve as a fast approximate alternative in simple or low dimensional scenarios, permutation testing remains necessary for reliable finite sample inference in general multivariate, since the dimension of latent position is unobserved in our framework. 

\graphicspath{{figs/}}

    \graphicspath{{figs/}}

\begin{figure}
        		\scriptsize
        		\begin{center}
        			\begin{tabular}{c}
        				\includegraphics[width=1.0\textwidth]{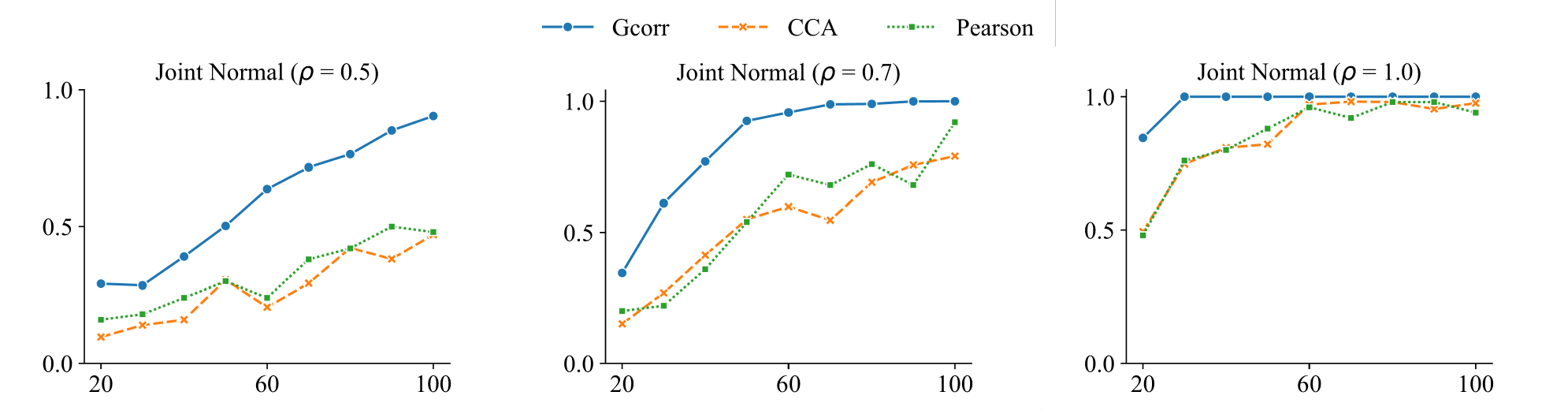}
        				\\
        			\end{tabular}
        		\end{center}
        		\caption{Comparing the testing power of Gcorr, CCA and Pearson under joint normal distribution.}
        		\label{fig:compare_JN_pow}
\end{figure}

\begin{figure}[htbp]
            \centering
            \begin{subfigure}[t]{0.3\textwidth}
                \centering
                \includegraphics[width=\linewidth]{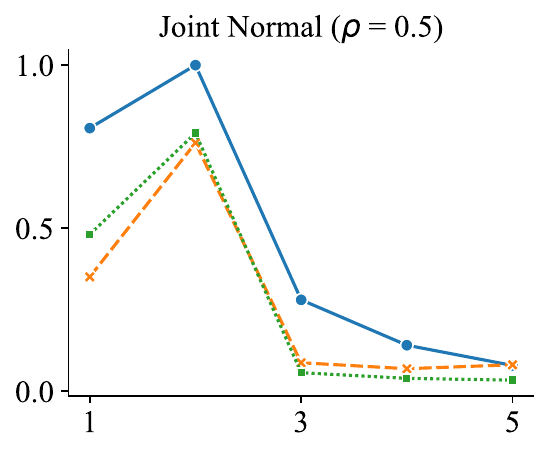}
                \caption{$n=100$}
                \label{fig:sub1}
            \end{subfigure}\hfill
            \begin{subfigure}[t]{0.3\textwidth}
                \centering
                \includegraphics[width=\linewidth]{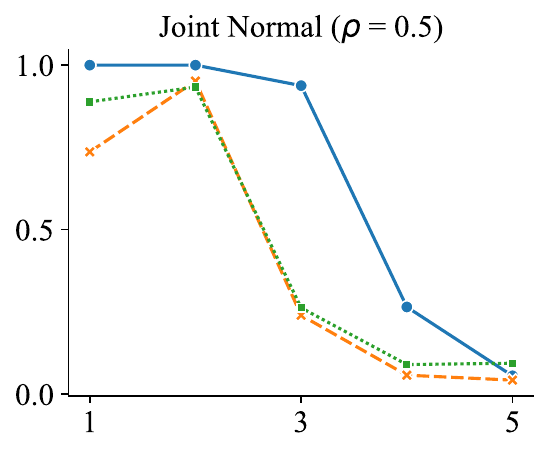}
                \caption{$n=200$}
                \label{fig:sub2}
            \end{subfigure}\hfill
            \begin{subfigure}[t]{0.3\textwidth}
                \centering
                \includegraphics[width=\linewidth]{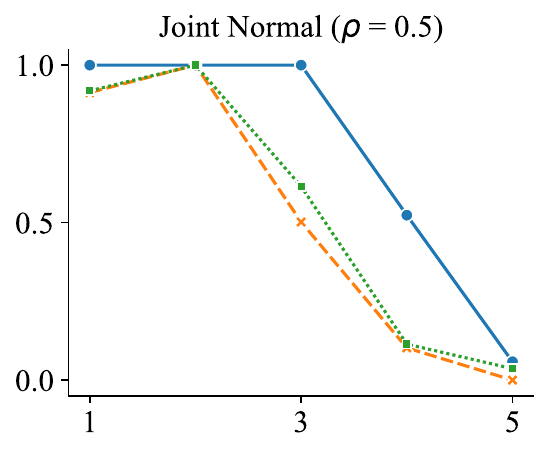}
                \caption{$n=300$}
                \label{fig:sub3}
            \end{subfigure}
            \caption{Comparing the testing power of Gcorr, MGC and Dcorr under joint normal distribution of different node size.}
            \label{fig:JN_pow_diff_node}
\end{figure}

We have conducted additional experiments comparing our method with CCA and Pearson correlation tests applied to graph-based embeddings derived from the Laplacian matrix, using a one-dimensional embedding for comparability. The proposed gCorr method exhibits consistently higher power than these CCA and Pearson tests; the corresponding results are reported in Figure~\ref{fig:compare_JN_pow}. In addition, we investigate how sample size affects testing power in a multivariate setting. As shown in Figure~\ref{fig:JN_pow_diff_node}, increasing the number of nodes leads to substantial power gains, which is consistent with theoretical expectations.


\graphicspath{{figs/}}

\graphicspath{{figs/}}

\bibliographystyle{chicago}

\bibliography{Gcorr_ref}

\begin{thebibliography}{}

\bibitem[\protect\citeauthoryear{Belkin and Niyogi}{Belkin and Niyogi}{2003}]{belkin2003laplacian}
Belkin, M. and P.~Niyogi (2003).
\newblock Laplacian eigenmaps for dimensionality reduction and data representation.
\newblock {\em Neural Computation\/}~{\em 15\/}(6), 1373--1396.

\bibitem[\protect\citeauthoryear{Bollob{\'a}s, Janson, and Riordan}{Bollob{\'a}s et~al.}{2007}]{bollobas2007phase}
Bollob{\'a}s, B., S.~Janson, and O.~Riordan (2007).
\newblock The phase transition in inhomogeneous random graphs.
\newblock {\em Random Structures and Algorithms\/}~{\em 31\/}(1), 3--122.

\bibitem[\protect\citeauthoryear{Chen, Hall, and Chklovskii}{Chen et~al.}{2006}]{chen2006wiring}
Chen, B.~L., D.~H. Hall, and D.~B. Chklovskii (2006).
\newblock Wiring optimization can relate neuronal structure and function.
\newblock {\em Proceedings of the National Academy of Sciences\/}~{\em 103\/}(12), 4723--4728.

\bibitem[\protect\citeauthoryear{Chu and Chen}{Chu and Chen}{2008}]{chu2008construction}
Chu, L.-H. and B.-S. Chen (2008).
\newblock Construction of a cancer-perturbed protein-protein interaction network for discovery of apoptosis drug targets.
\newblock {\em BMC systems biology\/}~{\em 2\/}(1), 1--17.

\bibitem[\protect\citeauthoryear{Coifman and Lafon}{Coifman and Lafon}{2006}]{coifman2006diffusion}
Coifman, R.~R. and S.~Lafon (2006).
\newblock Diffusion maps.
\newblock {\em Applied and Computational Harmonic Analysis\/}~{\em 21\/}(1), 5--30.

\bibitem[\protect\citeauthoryear{Cucker and Smale}{Cucker and Smale}{2002}]{cucker2002mathematical}
Cucker, F. and S.~Smale (2002).
\newblock On the mathematical foundations of learning.
\newblock {\em Bulletin of the American mathematical society\/}~{\em 39\/}(1), 1--49.

\bibitem[\protect\citeauthoryear{Cuturi}{Cuturi}{2013}]{cuturi2013sinkhorn}
Cuturi, M. (2013).
\newblock Sinkhorn distances: lightspeed computation of optimal transport.
\newblock In {\em Advances in Neural Information Processing Systems (NIPS)}, pp.\  2292–2300.

\bibitem[\protect\citeauthoryear{Davis and Kahan}{Davis and Kahan}{1970}]{davis1970rotation}
Davis, C. and W.~M. Kahan (1970).
\newblock The rotation of eigenvectors by a perturbation. iii.
\newblock {\em SIAM Journal on Numerical Analysis\/}~{\em 7\/}(1), 1--46.

\bibitem[\protect\citeauthoryear{Diaconis and Janson}{Diaconis and Janson}{2007}]{diaconis2007graph}
Diaconis, P. and S.~Janson (2007).
\newblock Graph limits and exchangeable random graphs.
\newblock {\em arXiv preprint arXiv:0712.2749\/}.

\bibitem[\protect\citeauthoryear{Ding and Du}{Ding and Du}{2023}]{ding2023detection}
Ding, J. and H.~Du (2023).
\newblock Detection threshold for correlated erd{\H{o}}s-r{\'e}nyi graphs via densest subgraph.
\newblock {\em IEEE Transactions on Information Theory\/}~{\em 69\/}(8), 5289--5298.

\bibitem[\protect\citeauthoryear{Ding, Fei, and Wang}{Ding et~al.}{2025}]{ding2025efficiently}
Ding, J., Y.~Fei, and Y.~Wang (2025).
\newblock Efficiently matching random inhomogeneous graphs via degree profiles.
\newblock {\em Annals of Statistics\/}~{\em 53\/}(4), 1808--1832.

\bibitem[\protect\citeauthoryear{D{\o}rum, Aln{\ae}s, Kaufmann, Richard, Lund, T{\o}nnesen, Sneve, Mathiesen, Rustan, Gjertsen, et~al.}{D{\o}rum et~al.}{2016}]{dorum2016age}
D{\o}rum, E.~S., D.~Aln{\ae}s, T.~Kaufmann, G.~Richard, M.~J. Lund, S.~T{\o}nnesen, M.~H. Sneve, N.~C. Mathiesen, {\O}.~G. Rustan, {\O}.~Gjertsen, et~al. (2016).
\newblock Age-related differences in brain network activation and co-activation during multiple object tracking.
\newblock {\em Brain and Behavior\/}~{\em 6\/}(11), e00533.

\bibitem[\protect\citeauthoryear{Fishkind, Sussman, Tang, Vogelstein, and Priebe}{Fishkind et~al.}{2013}]{fishkind2013consistent}
Fishkind, D.~E., D.~L. Sussman, M.~Tang, J.~T. Vogelstein, and C.~E. Priebe (2013).
\newblock Consistent adjacency-spectral partitioning for the stochastic block model when the model parameters are unknown.
\newblock {\em SIAM Journal on Matrix Analysis and Applications\/}~{\em 34\/}(1), 23--39.

\bibitem[\protect\citeauthoryear{Fosdick and Hoff}{Fosdick and Hoff}{2015}]{fosdick2015testing}
Fosdick, B.~K. and P.~D. Hoff (2015).
\newblock Testing and modeling dependencies between a network and nodal attributes.
\newblock {\em Journal of the American Statistical Association\/}~{\em 110\/}(511), 1047--1056.

\bibitem[\protect\citeauthoryear{Fukumizu, Gretton, Sun, and Sch\"{o}lkopf}{Fukumizu et~al.}{2007}]{fukumizu2007kernel}
Fukumizu, K., A.~Gretton, X.~Sun, and B.~Sch\"{o}lkopf (2007).
\newblock Kernel measures of conditional dependence.
\newblock In {\em Advances in Neural Information Processing Systems}, Volume~20.

\bibitem[\protect\citeauthoryear{Gjorgjieva, Biron, and Haspel}{Gjorgjieva et~al.}{2014}]{gjorgjieva2014neurobiology}
Gjorgjieva, J., D.~Biron, and G.~Haspel (2014).
\newblock Neurobiology of caenorhabditis elegans locomotion: where do we stand?
\newblock {\em Bioscience\/}~{\em 64\/}(6), 476--486.

\bibitem[\protect\citeauthoryear{Goodman and Sengupta}{Goodman and Sengupta}{2019}]{goodman2019caenorhabditis}
Goodman, M.~B. and P.~Sengupta (2019).
\newblock How caenorhabditis elegans senses mechanical stress, temperature, and other physical stimuli.
\newblock {\em Genetics\/}~{\em 212\/}(1), 25--51.

\bibitem[\protect\citeauthoryear{Gretton, Bousquet, Smola, and Sch{\"o}lkopf}{Gretton et~al.}{2005}]{gretton2005measuring}
Gretton, A., O.~Bousquet, A.~Smola, and B.~Sch{\"o}lkopf (2005).
\newblock Measuring statistical dependence with {H}ilbert-{S}chmidt norms.
\newblock In {\em International Conference on Algorithmic Learning Theory}, pp.\  63--77.

\bibitem[\protect\citeauthoryear{Gretton, Fukumizu, Teo, Song, Sch\"{o}lkopf, and Smola}{Gretton et~al.}{2007}]{gretton2007kernel}
Gretton, A., K.~Fukumizu, C.~Teo, L.~Song, B.~Sch\"{o}lkopf, and A.~Smola (2007).
\newblock A kernel statistical test of independence.
\newblock In {\em Advances in Neural Information Processing Systems}, Volume~20.

\bibitem[\protect\citeauthoryear{Hoff, Raftery, and Handcock}{Hoff et~al.}{2002}]{hoff2002latent}
Hoff, P.~D., A.~E. Raftery, and M.~S. Handcock (2002).
\newblock Latent space approaches to social network analysis.
\newblock {\em Journal of the American Statistical Association\/}~{\em 97\/}(460), 1090--1098.

\bibitem[\protect\citeauthoryear{Ito, Chiba, Ozawa, Yoshida, Hattori, and Sakaki}{Ito et~al.}{2001}]{ito2001comprehensive}
Ito, T., T.~Chiba, R.~Ozawa, M.~Yoshida, M.~Hattori, and Y.~Sakaki (2001).
\newblock A comprehensive two-hybrid analysis to explore the yeast protein interactome.
\newblock {\em Proceedings of the National Academy of Sciences\/}~{\em 98\/}(8), 4569--4574.

\bibitem[\protect\citeauthoryear{Janes, Nickerson, Frederick, and Kaufman}{Janes et~al.}{2012}]{janes2012prefrontal}
Janes, A.~C., L.~D. Nickerson, B.~d. Frederick, and M.~J. Kaufman (2012).
\newblock Prefrontal and limbic resting state brain network functional connectivity differs between nicotine-dependent smokers and non-smoking controls.
\newblock {\em Drug and Alcohol Dependence\/}~{\em 125\/}(3), 252--259.

\bibitem[\protect\citeauthoryear{Kalinke and Szab{\'o}}{Kalinke and Szab{\'o}}{2024}]{kalinke2024minimax}
Kalinke, F. and Z.~Szab{\'o} (2024).
\newblock The minimax rate of hsic estimation for translation-invariant kernels.
\newblock In {\em Advances in Neural Information Processing Systems (NIPS)}, pp.\  108468--108489.

\bibitem[\protect\citeauthoryear{Kato}{Kato}{1973}]{tosio1973continuity}
Kato, T. (1973).
\newblock Continuity of the map $s \to \left|s\right|$ for linear operators.
\newblock {\em Proceedings of the Japan Academy\/}~{\em 49\/}(3), 157 -- 160.

\bibitem[\protect\citeauthoryear{Klipp, Wade, and Kummer}{Klipp et~al.}{2010}]{klipp2010biochemical}
Klipp, E., R.~C. Wade, and U.~Kummer (2010).
\newblock Biochemical network-based drug-target prediction.
\newblock {\em Current Opinion in Biotechnology\/}~{\em 21\/}(4), 511--516.

\bibitem[\protect\citeauthoryear{Kloster and Gleich}{Kloster and Gleich}{2014}]{kloster2014heat}
Kloster, K. and D.~F. Gleich (2014).
\newblock Heat kernel based community detection.
\newblock In {\em Proceedings of the 20th ACM SIGKDD international conference on Knowledge discovery and data mining}, pp.\  1386--1395.

\bibitem[\protect\citeauthoryear{Kuhn, Campillos, Gonz{\'a}lez, Jensen, and Bork}{Kuhn et~al.}{2008}]{kuhn2008large}
Kuhn, M., M.~Campillos, P.~Gonz{\'a}lez, L.~J. Jensen, and P.~Bork (2008).
\newblock Large-scale prediction of drug-target relationships.
\newblock {\em FEBS Letters\/}~{\em 582\/}(8), 1283--1290.

\bibitem[\protect\citeauthoryear{Lee, Shen, Priebe, and Vogelstein}{Lee et~al.}{2019}]{lee2019network}
Lee, Y., C.~Shen, C.~E. Priebe, and J.~T. Vogelstein (2019).
\newblock Network dependence testing via diffusion maps and distance-based correlations.
\newblock {\em Biometrika\/}~{\em 106\/}(4), 857--873.

\bibitem[\protect\citeauthoryear{Levin, Roosta, Tang, Mahoney, and Priebe}{Levin et~al.}{2021}]{levin2021limit}
Levin, K.~D., F.~Roosta, M.~Tang, M.~W. Mahoney, and C.~E. Priebe (2021).
\newblock Limit theorems for out-of-sample extensions of the adjacency and {L}aplacian spectral embeddings.
\newblock {\em Journal of Machine Learning Research\/}~{\em 22\/}(194), 1--59.

\bibitem[\protect\citeauthoryear{Liu and Tse}{Liu and Tse}{2012}]{liu2012dynamics}
Liu, X.~F. and C.~Tse (2012).
\newblock Dynamics of network of global stock market.
\newblock {\em Accounting and Finance Research\/}~{\em 1\/}(2), 1--12.

\bibitem[\protect\citeauthoryear{Lyzinski, Fishkind, Fiori, Vogelstein, Priebe, and Sapiro}{Lyzinski et~al.}{2015}]{lyzinski2015graph}
Lyzinski, V., D.~E. Fishkind, M.~Fiori, J.~T. Vogelstein, C.~E. Priebe, and G.~Sapiro (2015).
\newblock Graph matching: Relax at your own risk.
\newblock {\em IEEE Transactions on Pattern Analysis and Machine Intelligence\/}~{\em 38\/}(1), 60--73.

\bibitem[\protect\citeauthoryear{Lyzinski, Fishkind, and Priebe}{Lyzinski et~al.}{2014}]{lyzinski2014seeded}
Lyzinski, V., D.~E. Fishkind, and C.~E. Priebe (2014).
\newblock Seeded graph matching for correlated {E}rd{\"o}s-{R}{\'e}nyi graphs.
\newblock {\em Journal of Machine Learning Research\/}~{\em 15\/}(1), 3513--3540.

\bibitem[\protect\citeauthoryear{Nichols and Holmes}{Nichols and Holmes}{2002}]{nichols2002nonparametric}
Nichols, T.~E. and A.~P. Holmes (2002).
\newblock Nonparametric permutation tests for functional neuroimaging: a primer with examples.
\newblock {\em Human Brain Mapping\/}~{\em 15\/}(1), 1--25.

\bibitem[\protect\citeauthoryear{Oliveira}{Oliveira}{2009}]{oliveira2009concentration}
Oliveira, R.~I. (2009).
\newblock Concentration of the adjacency matrix and of the laplacian in random graphs with independent edges.
\newblock {\em arXiv preprint arXiv:0911.0600\/}.

\bibitem[\protect\citeauthoryear{Rudi, Camoriano, and Rosasco}{Rudi et~al.}{2015}]{rudi2015less}
Rudi, A., R.~Camoriano, and L.~Rosasco (2015).
\newblock Less is more: Nystr{\"o}m computational regularization.
\newblock In {\em Advances in Neural Information Processing Systems (NIPS)}, pp.\  1657--1665.

\bibitem[\protect\citeauthoryear{Schafer}{Schafer}{2014}]{schafer2014mechanosensory}
Schafer, W. (2014).
\newblock Mechanosensory molecules and circuits in c. elegans.
\newblock {\em Pflugers Archiv: European Journal of Physiology\/}~{\em 467\/}(1), 39--48.

\bibitem[\protect\citeauthoryear{Schiavo, Reyes, and Fagiolo}{Schiavo et~al.}{2010}]{schiavo2010international}
Schiavo, S., J.~Reyes, and G.~Fagiolo (2010).
\newblock International trade and financial integration: a weighted network analysis.
\newblock {\em Quantitative Finance\/}~{\em 10\/}(4), 389--399.

\bibitem[\protect\citeauthoryear{Shen, Panda, and Vogelstein}{Shen et~al.}{2022}]{shen2022chi}
Shen, C., S.~Panda, and J.~T. Vogelstein (2022).
\newblock The chi-square test of distance correlation.
\newblock {\em Journal of Computational and Graphical Statistics\/}~{\em 31\/}(1), 254--262.

\bibitem[\protect\citeauthoryear{Shen, Priebe, and Vogelstein}{Shen et~al.}{2020}]{shen2020distance}
Shen, C., C.~E. Priebe, and J.~T. Vogelstein (2020).
\newblock From distance correlation to multiscale graph correlation.
\newblock {\em Journal of the American Statistical Association\/}~{\em 115\/}(529), 280--291.

\bibitem[\protect\citeauthoryear{Shi, Belkin, and Yu}{Shi et~al.}{2009}]{shi2009data}
Shi, T., M.~Belkin, and B.~Yu (2009).
\newblock Data spectroscopy: Eigenspaces of convolution operators and clustering.
\newblock {\em Annals of Statistics\/}~{\em 37\/}(6B), 3960--3984.

\bibitem[\protect\citeauthoryear{Sjoerds, Stufflebeam, Veltman, Van~den Brink, Penninx, and Douw}{Sjoerds et~al.}{2017}]{sjoerds2017loss}
Sjoerds, Z., S.~M. Stufflebeam, D.~J. Veltman, W.~Van~den Brink, B.~W. Penninx, and L.~Douw (2017).
\newblock Loss of brain graph network efficiency in alcohol dependence.
\newblock {\em Addiction Biology\/}~{\em 22\/}(2), 523--534.

\bibitem[\protect\citeauthoryear{Song, Priebe, and Tang}{Song et~al.}{2023}]{song2023independence}
Song, Y., C.~E. Priebe, and M.~Tang (2023).
\newblock Independence testing for inhomogeneous random graphs.
\newblock {\em arXiv preprint arXiv:2304.09132\/}.

\bibitem[\protect\citeauthoryear{Sriperumbudur and Sterge}{Sriperumbudur and Sterge}{2022}]{sriperumbudur2022approximate}
Sriperumbudur, B. and N.~Sterge (2022).
\newblock Approximate kernel {PCA} using random features: Computational vs. statistical trade-off.
\newblock {\em Annals of Statistics\/}~{\em 50\/}(5), 2713 -- 2736.

\bibitem[\protect\citeauthoryear{Sterge, Sriperumbudur, Rosasco, and Rudi}{Sterge et~al.}{2020}]{sterge2020gain}
Sterge, N., B.~Sriperumbudur, L.~Rosasco, and A.~Rudi (2020).
\newblock Gain with no pain: efficiency of kernel-{PCA} by {N}ystr{\"o}m sampling.
\newblock In {\em International Conference on Artificial Intelligence and Statistics}, pp.\  3642--3652.

\bibitem[\protect\citeauthoryear{Sumner and Popovi{\'c}}{Sumner and Popovi{\'c}}{2004}]{sumner2004deformation}
Sumner, R.~W. and J.~Popovi{\'c} (2004).
\newblock Deformation transfer for triangle meshes.
\newblock {\em ACM Transactions on Graphics (TOG)\/}~{\em 23\/}(3), 399--405.

\bibitem[\protect\citeauthoryear{Sussman, Park, Priebe, and Lyzinski}{Sussman et~al.}{2019}]{sussman2019matched}
Sussman, D.~L., Y.~Park, C.~E. Priebe, and V.~Lyzinski (2019).
\newblock Matched filters for noisy induced subgraph detection.
\newblock {\em IEEE Transactions on Pattern Analysis and Machine Intelligence\/}~{\em 42\/}(11), 2887--2900.

\bibitem[\protect\citeauthoryear{Sz{\'e}kely and Rizzo}{Sz{\'e}kely and Rizzo}{2009}]{szekely2009brownian}
Sz{\'e}kely, G.~J. and M.~L. Rizzo (2009).
\newblock Brownian distance covariance.
\newblock {\em Annals of Applied Statistics\/}~{\em 3\/}(4), 1236--1265.

\bibitem[\protect\citeauthoryear{Sz{\'e}kely and Rizzo}{Sz{\'e}kely and Rizzo}{2014}]{szekely2014partial}
Sz{\'e}kely, G.~J. and M.~L. Rizzo (2014).
\newblock Partial distance correlation with methods for dissimilarities.
\newblock {\em Annals of Statistics\/}~{\em 42\/}(6), 2382--2412.

\bibitem[\protect\citeauthoryear{Sz{\'e}kely, Rizzo, and Bakirov}{Sz{\'e}kely et~al.}{2007}]{szekely2007measuring}
Sz{\'e}kely, G.~J., M.~L. Rizzo, and N.~K. Bakirov (2007).
\newblock Measuring and testing dependence by correlation of distances.
\newblock {\em Annals of Statistics\/}~{\em 35\/}(6), 2769--2794.

\bibitem[\protect\citeauthoryear{Tang, Athreya, Sussman, Lyzinski, and Priebe}{Tang et~al.}{2017}]{tang2017nonparametric}
Tang, M., A.~Athreya, D.~L. Sussman, V.~Lyzinski, and C.~E. Priebe (2017).
\newblock A nonparametric two-sample hypothesis testing problem for random graphs.
\newblock {\em Bernoulli\/}~{\em 23\/}(3), 1599--1630.

\bibitem[\protect\citeauthoryear{Tang and Priebe}{Tang and Priebe}{2018}]{tang2018limit}
Tang, M. and C.~E. Priebe (2018).
\newblock Limit theorems for eigenvectors of the normalized laplacian for random graphs.
\newblock {\em Annals of Statistics\/}~{\em 46\/}(5), 2360--2415.

\bibitem[\protect\citeauthoryear{Tang, Sussman, and Priebe}{Tang et~al.}{2013}]{tang2013universally}
Tang, M., D.~L. Sussman, and C.~E. Priebe (2013).
\newblock Universally consistent vertex classification for latent positions graphs.
\newblock {\em Annals of Statistics\/}~{\em 41\/}(3), 1406--1430.

\bibitem[\protect\citeauthoryear{Varshney, Chen, Paniagua, Hall, and Chklovskii}{Varshney et~al.}{2011}]{varshney2011structural}
Varshney, L.~R., B.~L. Chen, E.~Paniagua, D.~H. Hall, and D.~B. Chklovskii (2011).
\newblock Structural properties of the caenorhabditis elegans neuronal network.
\newblock {\em PLoS computational biology\/}~{\em 7\/}(2), e1001066.

\bibitem[\protect\citeauthoryear{Wu, Xu, and Yu}{Wu et~al.}{2023}]{wu2023testing}
Wu, Y., J.~Xu, and S.~H. Yu (2023).
\newblock Testing correlation of unlabeled random graphs.
\newblock {\em Annals of Applied Probability\/}~{\em 33\/}(4), 2519--2558.

\bibitem[\protect\citeauthoryear{Xiong, Shen, Arroyo, and Vogelstein}{Xiong et~al.}{2019}]{xiong2019graph}
Xiong, J., C.~Shen, J.~Arroyo, and J.~T. Vogelstein (2019).
\newblock Graph independence testing.
\newblock {\em arXiv preprint arXiv:1906.03661\/}.

\bibitem[\protect\citeauthoryear{Young and Scheinerman}{Young and Scheinerman}{2007}]{young2007random}
Young, S.~J. and E.~R. Scheinerman (2007).
\newblock Random dot product graph models for social networks.
\newblock In {\em International Workshop on Algorithms and Models for the Web-Graph}, pp.\  138--149. Springer.

\end{thebibliography}
\end{document}